\documentclass[letterpaper,twocolumn,superscriptaddress,floatfix]{revtex4-2}
\usepackage{threeparttable}
\usepackage[utf8]{inputenc}
\usepackage[T1]{fontenc}
\usepackage{booktabs} 
\usepackage{inputenc}
\usepackage{bm}
\usepackage{multirow,amssymb,amsbsy,amsmath}
\usepackage{graphicx}
\usepackage{float}
\usepackage{verbatim}
\usepackage{color}
\makeatletter
\usepackage{pifont}
\usepackage{upgreek}
\usepackage{color}
\usepackage{indentfirst}
\usepackage{soul}
\usepackage{braket}
\usepackage[colorlinks=true,linkcolor=blue,citecolor=blue,urlcolor=blue]{hyperref}

\newcommand{\Rmnum}[1]{\expandafter\@slowromancap\romannumeral #1@}
\usepackage{etoolbox}
\usepackage{ragged2e}
\makeatletter
\apptocmd{\@caption}{\justifying}{}{}
\makeatother

\makeatletter
\renewcommand{\section}{\@startsection{section}{1}{0mm}
	{-\baselineskip}{0.5\baselineskip}{\bf\leftline}}
\makeatother

\begin{document}

\title{High-yield engineering and identification of oxygen-related modified divacancies in 4H-SiC}

\author{Qi-Cheng Hu}
\affiliation{Laboratory of Quantum Information, University of Science and Technology of China, Hefei, Anhui 230026, China}
\affiliation{Anhui Province Key Laboratory of Quantum Network, University of Science and Technology of China, Hefei, Anhui 230026, China}
\affiliation{CAS Center For Excellence in Quantum Information and Quantum Physics, University of Science and Technology of China, Hefei, Anhui 230026, China}
\affiliation{Hefei National Laboratory, University of Science and Technology of China, Hefei, Anhui 230088, China}

\author{Ji-Yang Zhou}
\affiliation{Laboratory of Quantum Information, University of Science and Technology of China, Hefei, Anhui 230026, China}
\affiliation{Anhui Province Key Laboratory of Quantum Network, University of Science and Technology of China, Hefei, Anhui 230026, China}
\affiliation{CAS Center For Excellence in Quantum Information and Quantum Physics, University of Science and Technology of China, Hefei, Anhui 230026, China}

\author{Shuo Ren}
\affiliation{Laboratory of Quantum Information, University of Science and Technology of China, Hefei, Anhui 230026, China}
\affiliation{Anhui Province Key Laboratory of Quantum Network, University of Science and Technology of China, Hefei, Anhui 230026, China}
\affiliation{CAS Center For Excellence in Quantum Information and Quantum Physics, University of Science and Technology of China, Hefei, Anhui 230026, China}

\author{Zhen-Xuan He}
\affiliation{Laboratory of Quantum Information, University of Science and Technology of China, Hefei, Anhui 230026, China}
\affiliation{Anhui Province Key Laboratory of Quantum Network, University of Science and Technology of China, Hefei, Anhui 230026, China}
\affiliation{CAS Center For Excellence in Quantum Information and Quantum Physics, University of Science and Technology of China, Hefei, Anhui 230026, China}
\affiliation{Hefei National Laboratory, University of Science and Technology of China, Hefei, Anhui 230088, China}

\author{Zhi-He Hao}
\affiliation{Laboratory of Quantum Information, University of Science and Technology of China, Hefei, Anhui 230026, China}
\affiliation{Anhui Province Key Laboratory of Quantum Network, University of Science and Technology of China, Hefei, Anhui 230026, China}
\affiliation{CAS Center For Excellence in Quantum Information and Quantum Physics,
University of Science and Technology of China, Hefei, Anhui 230026, China}
   
\author{Rui-Jian Liang}
\affiliation{Laboratory of Quantum Information, University of Science and Technology of China, Hefei, Anhui 230026, China}
\affiliation{Anhui Province Key Laboratory of Quantum Network, University of Science and Technology of China, Hefei, Anhui 230026, China}
\affiliation{CAS Center For Excellence in Quantum Information and Quantum Physics, University of Science and Technology of China, Hefei, Anhui 230026, China}
	
\author{Wu-Xi Lin}
\affiliation{Laboratory of Quantum Information, University of Science and Technology of China, Hefei, Anhui 230026, China}
\affiliation{Anhui Province Key Laboratory of Quantum Network, University of Science and Technology of China, Hefei, Anhui 230026, China}
\affiliation{CAS Center For Excellence in Quantum Information and Quantum Physics, University of Science and Technology of China, Hefei, Anhui 230026, China}
\affiliation{Hefei National Laboratory, University of Science and Technology of China, Hefei, Anhui 230088, China}	

\author{Xiangru Han}
\affiliation{HUN-REN Wigner Research Centre for Physics, Institute for Solid State Physics and Optics, P.O.\ Box 49, H-1525 Budapest, Hungary}

\author{Adam Gali}
\affiliation{HUN-REN Wigner Research Centre for Physics, Institute for Solid State Physics and Optics, P.O.\ Box 49, H-1525 Budapest, Hungary}
\affiliation{Department of Atomic Physics, Institute of Physics, Budapest University of Technology and Economics, M\H{u}egyetem rakpart 3., H-1111 Budapest, Hungary}
\affiliation{MTA-WFK "Lend\"ulet" Momentum Semiconductor Nanostructures Research Group, P.O.\ Box 49, H-1525 Budapest, Hungary}
   
\author{Jin-Shi Xu}
\altaffiliation{Email: jsxu@ustc.edu.cn}
\affiliation{Laboratory of Quantum Information, University of Science and Technology of China, Hefei, Anhui 230026, China}
\affiliation{Anhui Province Key Laboratory of Quantum Network, University of Science and Technology of China, Hefei, Anhui 230026, China}
\affiliation{CAS Center For Excellence in Quantum Information and Quantum Physics, University of Science and Technology of China, Hefei, Anhui 230026, China}
\affiliation{Hefei National Laboratory, University of Science and Technology of China, Hefei, Anhui 230088, China}

\author{Chuan-Feng Li}
\altaffiliation{Email: cfli@ustc.edu.cn}
\affiliation{Laboratory of Quantum Information, University of Science and Technology of China, Hefei, Anhui 230026, China}
\affiliation{Anhui Province Key Laboratory of Quantum Network, University of Science and Technology of China, Hefei, Anhui 230026, China}
\affiliation{CAS Center For Excellence in Quantum Information and Quantum Physics, University of Science and Technology of China, Hefei, Anhui 230026, China}
\affiliation{Hefei National Laboratory, University of Science and Technology of China, Hefei, Anhui 230088, China}
 
\author{Guang-Can Guo}
\affiliation{Laboratory of Quantum Information, University of Science and Technology of China, Hefei, Anhui 230026, China}
\affiliation{Anhui Province Key Laboratory of Quantum Network, University of Science and Technology of China, Hefei, Anhui 230026, China}
\affiliation{CAS Center For Excellence in Quantum Information and Quantum Physics, University of Science and Technology of China, Hefei, Anhui 230026, China}
\affiliation{Hefei National Laboratory, University of Science and Technology of China, Hefei, Anhui 230088, China}

\begin{abstract}

Modified divacancies in the 4H polytype of silicon carbide (SiC) exhibit enhanced charge stability and spin addressability at room temperature, making them attractive for quantum applications. However, their low formation yield and lack of direct structural identification have hindered progress. Here, we demonstrate a controllable method for high-yield engineering and identification of oxygen-related modified divacancy color centers in 4H-SiC via oxygen-ion implantation. Based on their distinct optical and spin-resonance characteristics, we experimentally resolve four types of modified divacancies. Furthermore, by measuring isotope-resolved $^{17}$O hyperfine interactions, we identify them as the four crystallographic configurations of oxygen-vacancy (OV) complexes. Remarkably, single OV centers account for over 90\% of the total defect population and exhibit superior optical properties and spin coherence compared with defects created by conventional carbon or nitrogen implantation. We characterize the zero-phonon lines of these OV centers and reveal distinct temperature-dependent behavior in spin-readout contrast. By optimizing implantation dose and annealing temperature, we achieve high-density ensembles and observe Rabi-oscillation beating patterns associated with different orientations of basal-type defects. These results establish a high-yield route for scalable engineering of these four oxygen-related modified divacancies in 4H-SiC and clarify their atomic structure, opening new opportunities for solid-state quantum technologies.
\\
\\
\end{abstract}
\maketitle
  
\date{\today}

\section{Introduction}
Solid-state spin qubits in silicon carbide (SiC) have emerged as promising building blocks for scalable quantum technologies~\cite{Son2006Divacancy,koehl2011room,widmann2015coherent,christle2015isolated,niethammer2016vector,bracher2017selective,nagy2019high,bourassa2020entanglement,lukin20204h,yan2020room,wang2020coherent,wang2021robust,abraham2021nanotesla,li2022room,babin2022fabrication,wang2023magnetic,jiang2023quantum,He2024low,son2020developing}. Among these, the neutral divacancy is an ideal quantum platform due to its bright emission, long spin coherence, and high readout fidelity~\cite{koehl2011room,christle2015isolated,He2024low}. Experimental and theoretical studies have identified PL1--PL4 as carbon--silicon divacancy complexes, which correspond to adjacent missing atoms at cubic ($k$) and hexagonal ($h$) lattice sites~\cite{Son2006Divacancy,koehl2011room,falk2013polytype,gordon2015defects,magnusson2018excitation,davidsson2018first}.

Beyond these well-understood divacancies, another group of defects, labeled PL5--PL8, exhibits zero-field splitting (ZFS) and zero-phonon line (ZPL) characteristics similar to PL1--PL4 and has thus been tentatively classified as modified divacancies~\cite{yan2020room,li2022room,He2024low,falk2013polytype,son2022modified,shafizadeh2025evolution}. Among them, single PL5 and PL6 centers display strong room-temperature photoluminescence, stable charge states, and pronounced spin contrast~\cite{li2022room,He2024low}. However, despite their attractive quantum properties, both their low formation yield and the lack of direct microscopic structural identification have substantially hindered further progress. In addition, the absence of a distinct ZPL for PL7~\cite{falk2013polytype} further obscures the microscopic origin of this family of defects.

Early models attributed PL5--PL7 to stacking-fault regions~\cite{ivady2019stabilization}, but subsequent experiments have challenged this hypothesis. Specifically, irradiation-induced stacking faults in $n$-type 4H-SiC failed to generate PL5--PL7 centers~\cite{son2022modified}. These inconsistencies have motivated the search for alternative atomic configurations.

Recent theoretical investigations have identified oxygen-related defects in SiC as a distinct family of color centers~\cite{kobayashi2023oxygen,iwamoto2024oxygen}. Based on this insight, \textit{ab initio} calculations have proposed that neutral oxygen-substituted silicon vacancies at the $hk$ and $hh$ sites may account for PL5 and PL6~\cite{bai2025origin}. These models successfully reproduce the experimentally observed ZFS and ZPL values and predict symmetry axes consistent with experimental observations. However, direct experimental evidence for this structural assignment, particularly from isotope-resolved hyperfine spectroscopy, has remained unavailable.

Here, we present a detailed investigation of the color centers generated by oxygen-ion implantation in 4H-SiC. At a dose of $1\times10^{8}$~cm$^{-2}$, four dominant color centers are observed. Two of them exhibit optical and spin characteristics consistent with the previously identified PL5 and PL6 defects, while the other two—designated PL7$'$ and PL8$'$—display distinct properties. PL7$'$ is related to the previously reported basal-like PL7, sharing several spectral features, whereas PL8$'$ represents a newly identified $c$-axis–like defect; its zero-field splitting (ZFS) and zero-phonon line (ZPL) values agree well with theoretical predictions for oxygen–vacancy complexes~\cite{bai2025origin}. Notably, the generation yield of these single modified divacancies is exceptionally high—together they account for more than 90\% of the total defect population.

Room-temperature spin coherence measurements reveal prolonged $T_{2}$ times for PL5 and PL6 compared with carbon- and nitrogen-implanted samples. Temperature-dependent ODMR measurements reveal distinct contrast evolutions among the four signals: the contrast of the PL7$'$ resonance decreases at low temperature, that of PL8$'$ increases, the PL6 contrast remains almost unchanged, and the PL5 resonances become mixed—one branch maintains its contrast while the other shows a noticeable reduction. Ensemble measurements further reveal that high-dose oxygen implantation suppresses the PL1–PL4 emissions while enhancing the PL5–PL8$'$ features, resulting in the appearance of clear Rabi oscillation beating patterns associated with the basal defects at room temperature.

More importantly, isotope-resolved $^{17}$O hyperfine measurements allow us to directly identify the atomic structure of these four oxygen-related modified divacancies. Distinct hyperfine splittings are resolved for all four defect species, providing direct evidence that these four centers correspond to the four crystallographic configurations of O$_\text{C}$V$_\text{Si}$ complexes, in which an oxygen atom substitutes for a carbon site and a neighboring silicon vacancy is formed.

These results establish a high-yield route for engineering these four oxygen-related modified divacancies in 4H-SiC and clarify their microscopic structure. By linking the historically labeled modified-divacancy signals to structurally identified O$_\text{C}$V$_\text{Si}$ centers, this work provides a solid foundation for future applications of SiC spin defects in quantum sensing, quantum communication, and scalable solid-state quantum technologies.

\section{Results and Discussion}
\subsection{Preparation of single oxygen-related modified divacancies}

\begin{figure*}[htbp]
\centering
\includegraphics[scale = 0.55]{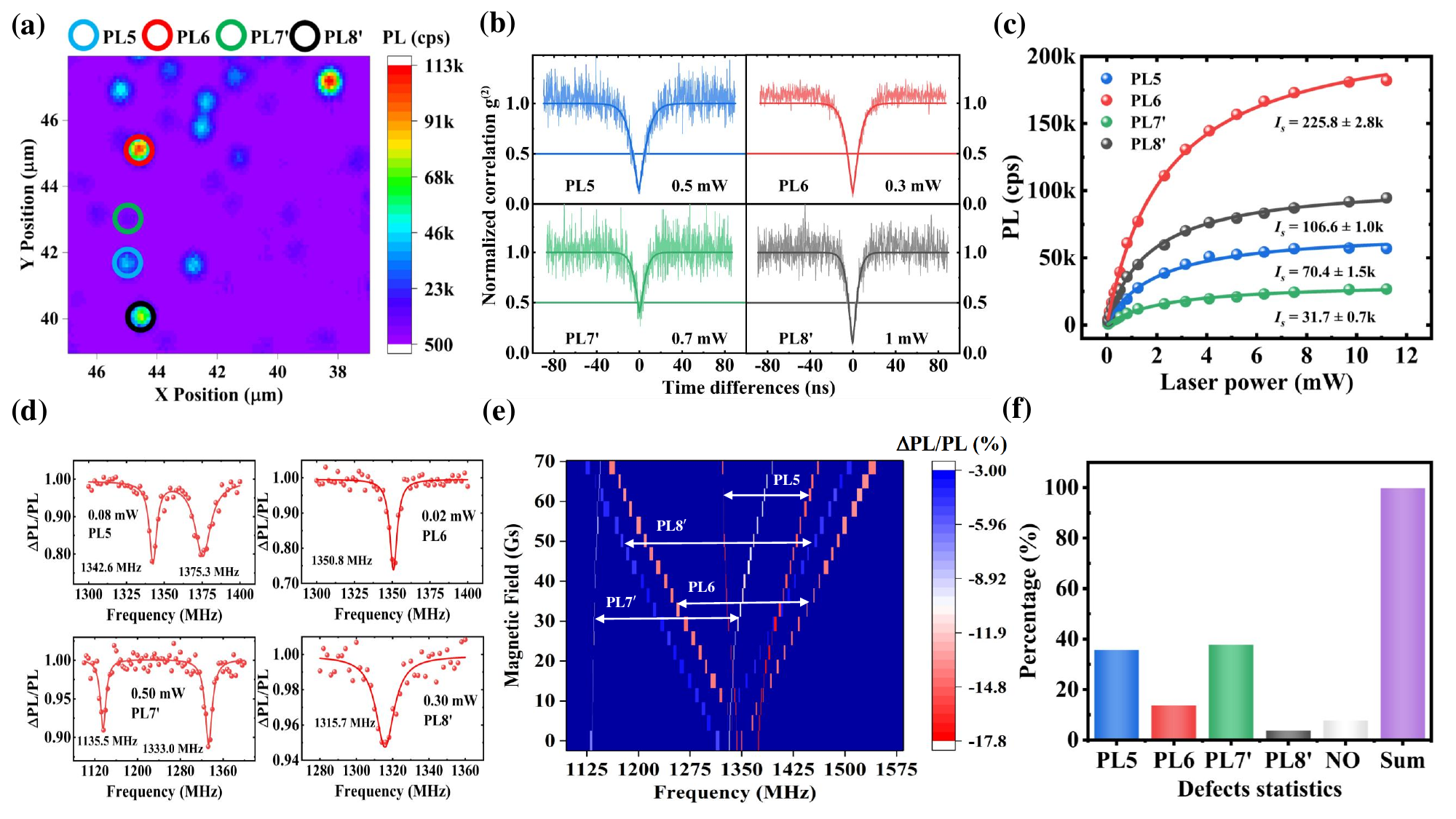}
\caption{Single {oxygen-related} modified divacancies in oxygen ion–implanted samples.
(a) Photoluminescence (PL) map of a 10 × 10 $\mu$m$^2$ region under 914 nm excitation with 2 mW laser power. The blue, red, green, and black circles indicate the locations of PL5, PL6, PL7$'$, and PL8$'$ color centers, respectively.
(b) Second-order autocorrelation functions $g^{(2)}(\tau)$ of PL5–PL8$'$, shown without background subtraction.
(c) Saturation curves of fluorescence count rates measured for PL5–PL8$'$.
(d) Zero-field optically detected magnetic resonance (ODMR) spectra of PL5–PL8$'$, recorded under different laser powers and -20 dB microwave power.
(e) Magnetic-field-dependent ODMR responses of PL5–PL8$'$ at 0.8 mW.
(f) Statistical distribution of color center types among 149 randomly selected emitters at room temperature.
}
\label{Figure 1}
\end{figure*}
The formation yield of modified divacancies is inherently low for carbon and nitrogen implantation, often requiring high doses to obtain isolated defects. However, excessive doses increase lattice damage and degrade the signal-to-noise ratio. To address this trade-off, we employed oxygen-ion implantation at a reduced dose of $1\times10^{8}$~cm$^{-2}$ to generate  single oxygen-related modified divacancy. Details of the sample preparation are provided in the \textit{Experimental Section}.

Fig.~\ref{Figure 1}a shows a $10\times10~\mu$m$^2$ photoluminescence (PL) map recorded under 2.0~mW excitation. Four types of spin-active color centers—PL5, PL6, PL7$'$, and PL8$'$—are identified and highlighted by colored circles. These emitters exhibit high signal-to-noise ratios without any post-annealing surface treatment. Details on the comparison of carbon, nitrogen, and oxygen ion implantation are provided in Supplementary Information A.

The single-photon nature of the emitters was verified by measuring the second-order autocorrelation function $g^{(2)}(\tau)$ as a function of time delay $\tau$ (Fig.~\ref{Figure 1}b). The experimental raw data were analyzed by fitting to the following model~\cite{castelletto2014silicon}:
$g^{(2)}(\tau) = 1 - (1 + a)e^{-|\tau - \tau_0|/t_1} + b e^{-|\tau - \tau_0|/t_2}$, where $\tau_0$ represents the time delay at zero point, and $t_1$, $t_2$, $a$, $b$ are fitting parameters. For all four defects, $g^{(2)}(0)$ values remain below 0.5 without background subtraction, confirming single-photon emission. Power-dependent $g^{(2)}(\tau)$ measurements are provided in Supplementary Information B.

The fluorescence saturation curves in Fig.~\ref{Figure 1}c yield saturation count rates and powers of $70.4\pm1.5$~kcps and $1.93 \pm 0.13$~mW for PL5, $225.8 \pm 2.8$~kcps and $2.36 \pm 0.09$~mW for PL6, $31.7 \pm 0.7$~kcps and $2.33 \pm 0.14$~mW for PL7$'$, and $106.6 \pm 1.0$~kcps and $1.66\pm0.05$~mW for PL8$'$. Among them, PL6 is the brightest single emitter at room temperature—consistent with results from carbon-ion implantation~\cite{li2022room}—whereas PL7$'$ is the weakest, being more than seven times dimmer than PL6.

Continuous-wave optically detected magnetic resonance (ODMR) spectra of individual centers are presented in Fig.~\ref{Figure 1}d. PL5 exhibits two zero-field resonances at 1342.6 and 1375.3~MHz with 20\% contrast; the extracted zero-field splitting parameters are $D = 1358.9$~MHz and $E = 16.6$~MHz. PL6 shows a single sharp resonance at 1350.8~MHz with 25\% contrast, consistent with previous reports~\cite{li2022room}. PL7$'$ displays a double-resonance feature at 1135.5 and 1333.0~MHz, with $D = 1234.2$~MHz and $E = 98.8$~MHz. In previous ensemble studies~\cite{falk2013polytype,shafizadeh2025evolution}, the left branch of PL3 and the PL7 center were reported at 1135 MHz and 1333 MHz, respectively. However, in our room-temperature single-defect measurements, we did not observe the peak-shaped right branch expected for PL3. Instead, we repeatedly detected a strong resonance at 1333.0 MHz together with its corresponding lower-frequency branch at 1135.5 MHz. We therefore assign PL7$'$ to the previously reported PL7 center, indicating that PL7 intrinsically possesses a double-resonant structure. PL8$'$ exhibits a single resonance at 1315.7~MHz with a lower $\sim$5\% contrast, different from PL1 (1323.5~MHz)~\cite{li2022room}, hinting at a new type of room-temperature spin qubit.

The magnetic-field-dependent ODMR spectra (Fig.~\ref{Figure 1}e) confirm that PL5 and PL7$'$ are basal-plane oriented, while PL6 and PL8$'$ align along the c-axis. PL5, PL6, and PL7$'$ display pronounced contrasts, whereas PL8$'$ retains a lower ($\sim$5\%) room-temperature contrast. Statistical analysis of 149 emitters (Fig.~\ref{Figure 1}f) reveals that approximately 92\% correspond to {oxygen-related} modified divacancy centers. Detailed statistical results are provided in Supplementary Information C. These observations strongly indicate that oxygen incorporation plays a crucial role in the formation of these color centers.

\subsection{Coherent control of single {oxygen-related} modified divacancies}
\begin{figure*}[htbp]
\centering
\includegraphics[scale = 0.55]{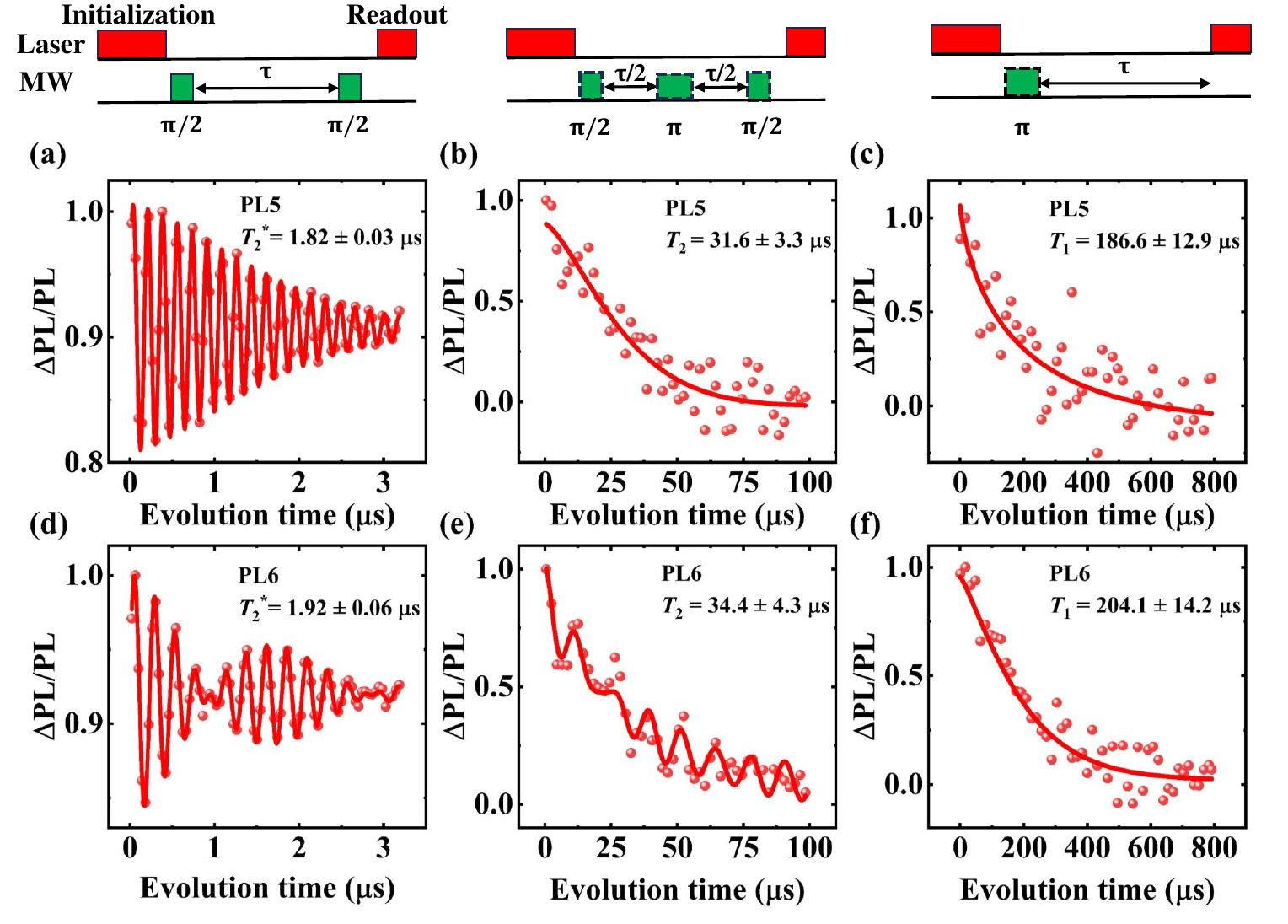}
  \caption{Spin coherent control of single PL5 and PL6 centers.
(a) Ramsey measurement of the inhomogeneous dephasing time ($T_{2}^{*}$) for PL5 without an external magnetic field.
(b) Spin coherence time ($T_{2}$) measurement of the left branch of PL5 without an external magnetic field.
(c) Spin-lattice relaxation time ($T_{1}$) measurement of PL5 without an external magnetic field.
(d) Ramsey measurement of the inhomogeneous dephasing time ($T_{2}^{*}$) for PL6 at 180 G.
(e) Spin coherence time ($T_{2}$) measurement of the left branch of PL6 at 180 G.
(f) Spin-lattice relaxation time ($T_{1}$) measurement of PL6 at 180 G.
The corresponding laser and microwave (MW) pulse sequences are shown above.} 
 \label{Figure 2}
\end{figure*}
Spin coherence properties are central to the applicability of color centers in quantum technologies. The modified divacancies represent a typical three-level system ($S=1$), whose spin Hamiltonian under an external magnetic field can be expressed as:
\begin{equation}
\mathcal{H_{MD}} = D \left( S_z^2 - \frac{S(S+1)}{3} \right) + E (S_x^2 - S_y^2) + g_{e} \mu_B \mathbf{B} \cdot \mathbf{S},
\label{electron}
\end{equation}
where $D$ and $E$ are the zero-field splitting parameters, $g_e$ is the electron $g$-factor, $\mu_B$ is the Bohr magneton, $\mathbf{B}$ is the applied magnetic field, and $\mathbf{S}$ is the spin operator. We characterized PL5, PL6, PL7$'$, and PL8$'$ in terms of their Ramsey inhomogeneous dephasing time ($T_2^*$), spin coherence time ($T_2$), and spin--lattice relaxation time ($T_1$), all measured under a relatively high laser power of 0.8~mW. The corresponding laser and microwave (MW) pulse sequences are shown schematically above the measured results.

The basal-oriented PL5 color center hosts mixed $m_{s}=\pm 1$ energy levels. With the left branch designated as the $\ket{+}=1/\sqrt{2}(\ket{1}+\ket{-1})$ state, optical polarization initializes the system into $\ket{0}$ state, enabling coherent manipulation from $\ket{0}$ to $\ket{+}$ via microwave excitation. By contrast, the c-axis oriented PL6 color center exhibits a lifting of the degeneracy of the $m_{s}=\pm 1$ levels under an applied magnetic field along the c-axis. With the left branch defined as the $\ket{1}$ state, we performed coherent measurements on the $\ket{0}$ to $\ket{1}$ transition.

For PL5 (Fig.~\ref{Figure 2}a--2c), we obtained $T_2^* = 1.82\pm0.03~\mu$s, $T_2 = 31.6\pm3.3~\mu$s, and $T_1 = 186.6\pm12.9~\mu$s without an external magnetic field. These values surpass those of carbon-implanted defects~\cite{li2022room}. PL6 (Fig.~\ref{Figure 2}d--2f) exhibited even longer coherence, with $T_2^* = 1.92\pm0.06~\mu$s, $T_2 = 33.4\pm4.3~\mu$s, and $T_1 = 204.1\pm14.2~\mu$s under a 180~G magnetic field along the c-axis. To further benchmark the coherence performance, we compared the $T_2$ times of PL6 centers generated by carbon, nitrogen, and oxygen ion implantation in the same pristine sample; the corresponding results are provided in Supplementary Information D. In the Ramsey fringes of PL6, we observed characteristic modulations arising from weakly coupled $^{29}$Si nuclear spins~\cite{li2022room}. The natural abundance of $^{29}$Si (4.7\%) and $^{13}$C (1.1\%) provides a dense nuclear spin reservoir, making these centers promising candidates for hybrid electron--nuclear quantum registers. The coherence data for PL7$'$ and PL8$'$ are further provided in Supplementary Information~E. 

Owing to their excellent optical and spin properties, particularly the controllable implantation depth, these single oxygen-related modified defects are highly suitable for quantum sensing~\cite{li2025non}. Stopping and Range of Ions in Matter (SRIM) simulations for the implantation process are presented in Supplementary Information {F}. Among them, PL6 is particularly attractive because of its superior ODMR contrast and brightness, as well as its stable charge state in the shallow region~\cite{He2024low}. Accordingly, we further demonstrate magnetic-field sensing using Gd$^{3+}$ spins based on shallow PL6 centers prepared by oxygen-ion implantation, as detailed in Supplementary Information G.

\subsection{Low-temperature properties of single oxygen-related modified divacancies}
\begin{figure*}[htbp]
\centering
\includegraphics[scale = 0.55]{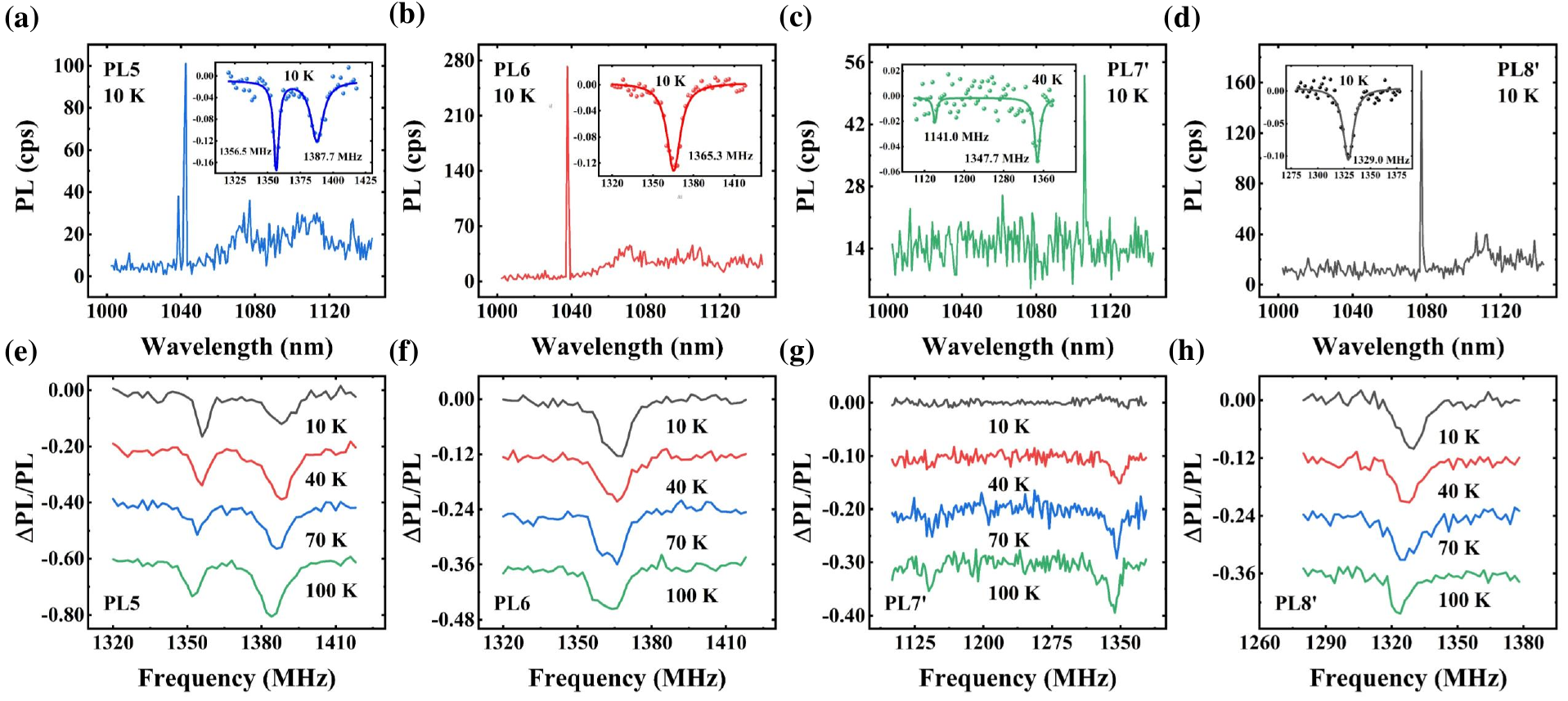}
\caption{Zero-phonon line (ZPL) and zero-field splitting (ZFS) of single {oxygen-related} modified divacancy defects.
(a-d) Measured zero-phonon line (ZPL) and zero-field splitting (ZFS) of a single emitter at cryogenic temperature.
(e–h) Temperature-dependent ODMR spectra of PL5–PL8$'$.
}
\label{Figure 3}
\end{figure*}
Cryogenic ODMR and PL spectroscopy (Fig.~\ref{Figure 3}a--3d) were performed on the same single centers tracked during cooling. PL5 exhibits a ZPL at 1042.2~nm and zero-field splitting (ZFS) of 1356.5 and 1387.7~MHz; PL6 shows a ZPL at 1037.8~nm and a ZFS of 1365.3~MHz, consistent with earlier reports~\cite{koehl2011room,li2022room,falk2013polytype,shafizadeh2025evolution}. {In addition,} a weak emission near the PL5 ZPL appears in most samples, reminiscent of the V1/V1$'$ doublet in silicon vacancies~\cite{nagy2018quantum}. 

PL7$'$ shows a ZPL at 1106.2~nm, close to that of PL3 (1108~nm)~\cite{koehl2011room,li2022room,falk2013polytype,magnusson2018excitation,shafizadeh2025evolution} and is comparable to the previously reported PL3a~\cite{shafizadeh2025evolution}. This spectral proximity could also explain why the PL7$'$/PL7 line was previously unresolved at low temperatures. PL8$'$ exhibits a ZPL at 1077.1~nm, near that of PL4 (1078~nm), but displays distinct ODMR features with c-axis orientation.

Temperature-dependent ODMR spectra reveal contrasting behaviors among the centers. For PL5 (Fig.~\ref{Figure 3}e), the right branch of the ODMR signal weakens markedly below 10~K, while the left branch remains stable. PL6 shows minimal change from 100~K to 10~K (Fig.~\ref{Figure 3}f). In contrast, the ODMR contrast of PL7$'$ vanishes near 10~K (Fig.~\ref{Figure 3}g). Conversely, PL8$'$ exhibits increasing ODMR contrast upon cooling, reaching nearly twice its room-temperature value at 10~K (Fig.~\ref{Figure 3}h). These trends highlight PL8$'$ as a promising spin qubit candidate for cryogenic quantum applications. The distinct temperature-dependent behaviors further confirm the differences among these defects and provide a pathway for investigating temperature-dependent intersystem crossing processes.

The measured ZPL energies of 1.1899, 1.1947, 1.1208, and 1.1513~eV for PL5, PL6, PL7$'$, and PL8$'$, respectively agree well with the theoretically calculated values of oxygen-vacancy complexes~\cite{bai2025origin}. Together with their distinct optical and spin signatures, this agreement supports the assignment of these four oxygen-related modified divacancies to the corresponding crystallographic configurations of O$_\text{C}$V$_\text{Si}$ complexes. The basic parameters of different centers, together with the schematic diagram of the oxygen-vacancy centers, are provided in Supplementary Information~H.

\subsection{High-concentration ensembles {of oxygen-related modified divacancies}}
 \begin{figure*}[htbp]
\centering
\includegraphics[scale = 0.5]{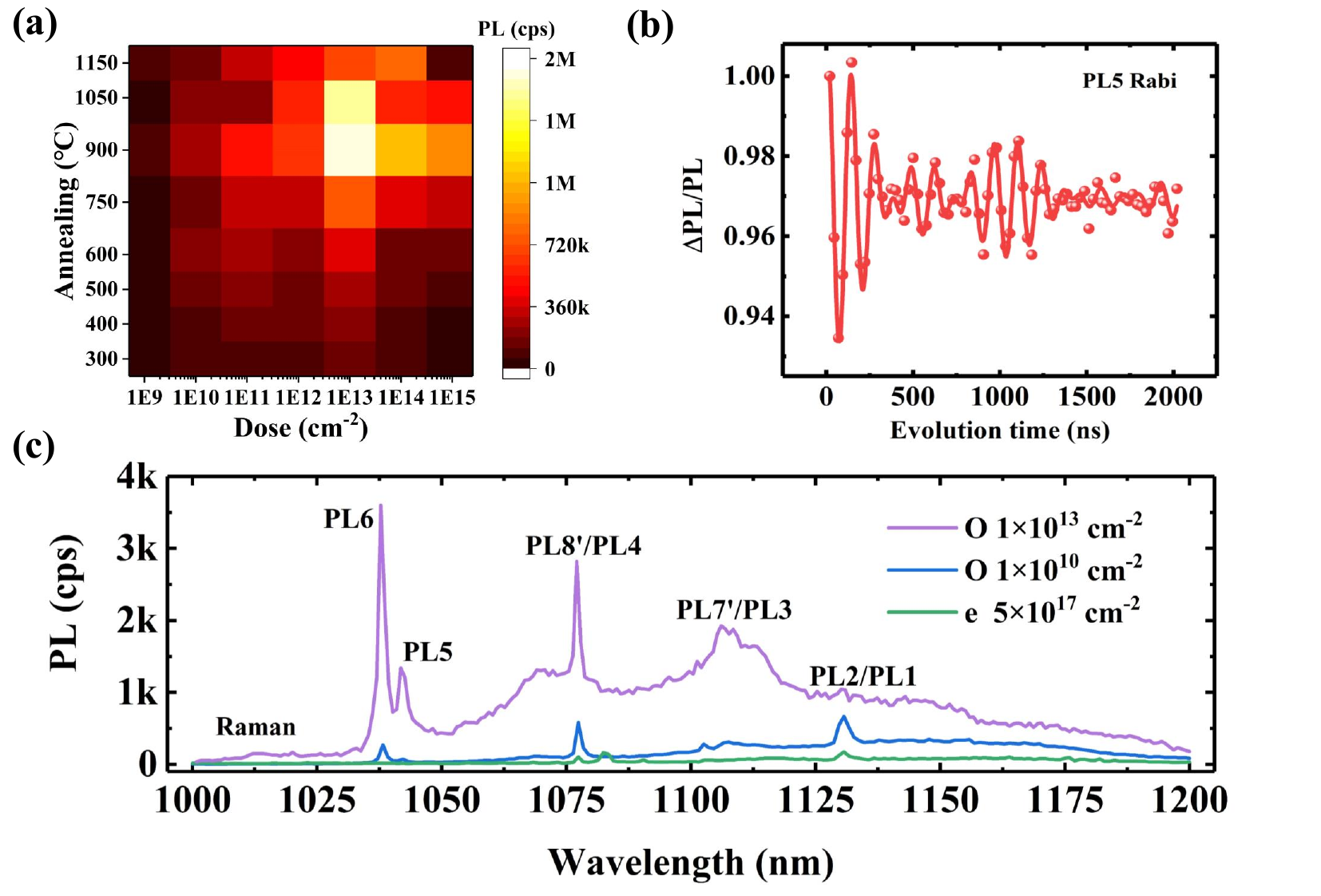}
\caption{Optical and spin properties of ensemble samples.
(a) Optimization of implantation dose and annealing temperature for oxygen-implanted ensembles. (b) Rabi oscillation measurements of PL5 centers with the characteristic beating patterns.
(c) Low-temperature ZPL spectra of ensembles with oxygen ion implantation of {different doses}, as well as electron irradiation.
}
\label{Figure 4}
\end{figure*}
We next optimized ensemble samples by systematically tuning the oxygen implantation dose and annealing temperature. The corresponding PL intensities are shown in Fig.~\ref{Figure 4}a. Under 0.1~mW excitation, samples implanted with $1\times10^{13}$~cm$^{-2}$ oxygen ions and annealed at 900$^{\circ}$C for 0.5~h in vacuum exhibited PL intensities exceeding 1.7~Mcps—significantly higher than those in previously reported ensemble samples~\cite{wang2020optimization}. The estimated defect concentration is approximately $1\times10^{16}$~cm$^{-3}$, which is promising for the development of quantum sensing technologies under extreme conditions, such as high-pressure magnetic-field detection~\cite{liu2022pressure}. The silicon vacancies in the same samples were optimized in Supplementary Information~I.

We further found that the nitrogen concentration strongly affected the observable fluorescence of these oxygen-related modified divacancies. When the nitrogen concentration increased from below $1\times10^{14}$~cm$^{-3}$ to $1\times10^{19}$~cm$^{-3}$, the PL intensity at the optimal implantation dose and annealing temperature decreased by about a factor of ten. This reduction may arise from decreased stability of the relevant emissive charge state in highly nitrogen-doped samples. A detailed investigation of the role of nitrogen concentration is presented in Supplementary Information J.

Rabi oscillation measurements of PL5 centers (Fig.~\ref{Figure 4}b) revealed characteristic beating patterns, consistent with previous reports using substrate samples~\cite{koehl2011room}. To fit the PL5 Rabi oscillations, we employed a triple-cosine function with an exponential decay envelope:

\begin{equation}
\begin{aligned}
f(t) = a &+ b \cdot \exp\left[-\left(\frac{t}{\tau}\right)^n\right]\times \prod_{i=1}^{3} \cos\left(2\pi\Omega_i t + \phi_i\right),
\end{aligned}
\end{equation}
where the observed beating pattern reflects the superposition of multiple Rabi frequencies under a single microwave driving field, indicating complex coherent dynamics in the ensemble system. This behavior arises from the anisotropic nature of PL5 centers, which possess three basal orientations within the lattice. Each orientation experiences a distinct microwave driving strength due to the vectorial nature of the microwave field, resulting in orientation-dependent Rabi frequencies ($\Omega_1$, $\Omega_2$ and $\Omega_3$) that manifest as observable beatings in the time-domain signal. The parameters $a$, $b$, $\tau$, $\phi_1$, $\phi_2$ and $\phi_3$ are fitting parameters. The optical and spin characteristics of the modified divacancies ensembles, presented in Supplementary Information K and Supplementary Information L, are consistent with those observed in the corresponding single-defect cases.

Comparative ensemble studies using different doses of oxygen ion implantation, as well as electron irradiation under the same annealing conditions, are summarized in Fig.~\ref{Figure 4}c. Electron-irradiated samples showed negligible ZPL emissions from modified divacancies. Increasing the oxygen dose from $1\times10^{10}$~cm$^{-2}$ to $1\times10^{13}$~cm$^{-2}$ suppressed PL1 and PL2 entirely, while PL6 became the dominant ZPL contributor. This trend indicates that oxygen ions promote a higher conversion yield {toward oxygen-related} modified divacancies. In Supplementary Information L, we further present temperature-dependent ODMR spectra for the sample implanted at a dose of $1\times10^{13}$~cm$^{-2}$, where no discernible signals from PL1--PL4 are observed. For comparison, we also examined ensemble spectra from carbon- and nitrogen-ion-implanted samples prepared at the same dose. Both the overall fluorescence intensity and the relative proportion of modified divacancies in those samples are markedly lower than in the oxygen-ion-implanted sample, as detailed in Supplementary Information M. These results indicate that, under optimized oxygen implantation conditions, oxygen-related modified divacancies are generated with high efficiency, and the strong fluorescence signal of the ensemble originates predominantly from this defect family.

\subsection{{Hyperfine coupling of $^{17}$O with oxygen-related modified divacancies}}

\begin{figure*}[htbp]
\centering
\includegraphics[scale = 0.90]{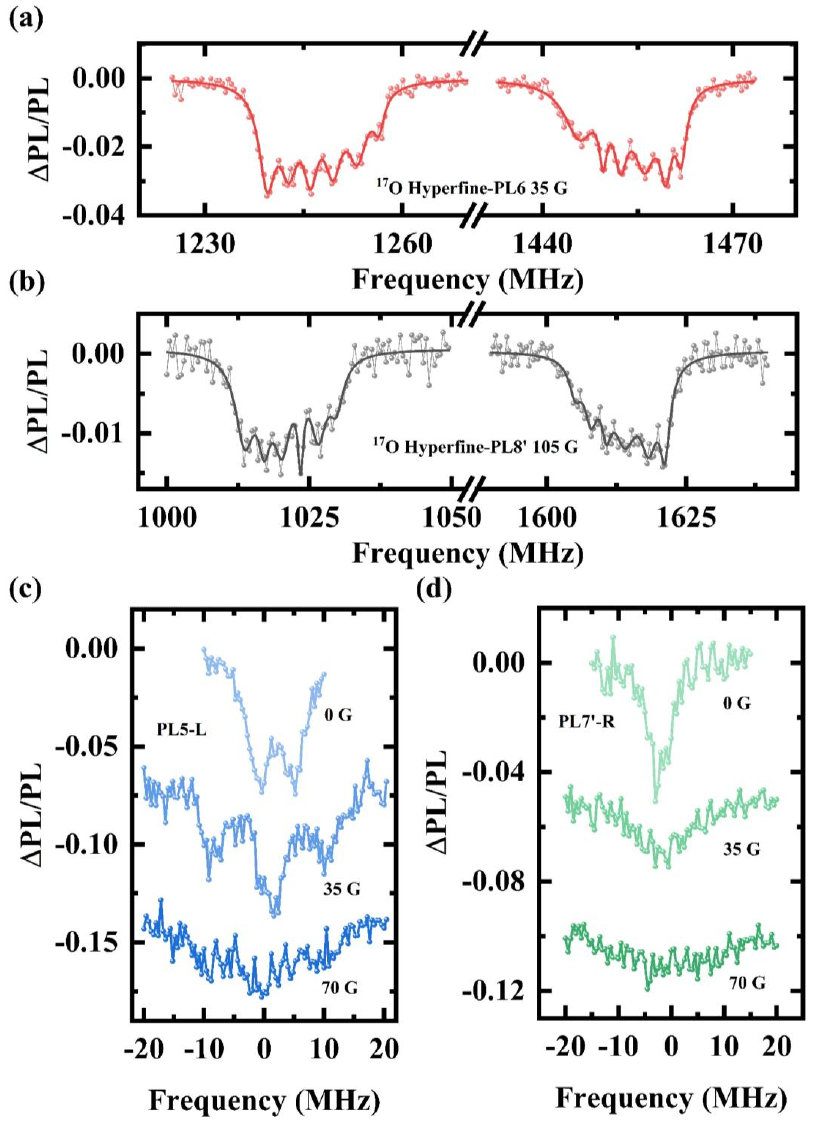}
\caption{Hyperfine interaction measurements for defects created by $^{17}$O implantation. (a) and (b) ODMR spectra of the c-axis-oriented PL6 and PL8$'$ centers under magnetic fields of 35 G and 105 G, applied along the crystal c-axis. The spectra of PL6 and PL8$'$ are fitted with six Lorentzian components. (c) and (d) ODMR spectra of the left branch of the basal-type PL5 center (PL5-L) and the right branch of the basal-type PL7$'$ center (PL7$'$-R), under external magnetic fields of varying strength applied along the c-axis.}
\label{Figure 5}
\end{figure*}

Hyperfine interactions with intrinsic nuclear spins provide one of the most direct and decisive probes for identifying the microscopic structure of color centers. To this end, we optimized the $^{17}$O ion-implantation process and obtained samples in which intrinsic $^{17}$O hyperfine coupling could be resolved. This observation provides direct evidence that oxygen is incorporated into the microscopic structure of these modified divacancies, thereby strongly supporting their assignment as oxygen-vacancy (OV) centers. Detailed results on the current discrimination of $^{17}$O ions are provided in Supplementary Information N.

The interaction Hamiltonian between the defect electron spin and the $^{17}$O nuclear spin can be written as~\cite{pezzagna2024polymorphs}
\begin{equation}
\mathcal{H} = \mathcal{H}_{MD} + \gamma_n^{(\mathrm{O})} \mathbf{B} \cdot \mathbf{I}^{(\mathrm{O})} + \mathcal{H}_Q + \mathbf{S} \cdot \mathbf{A}^{(\mathrm{O})} \cdot \mathbf{I}^{(\mathrm{O})},
\end{equation}
where $\mathcal{H}_{MD}$ denotes the electron-spin Hamiltonian of the modified divacancy, as defined in Eq.~(\ref{electron}); $\gamma_n^{(\mathrm{O})}$ is the gyromagnetic ratio of the $^{17}$O nucleus; $\mathbf{I}^{(\mathrm{O})}$ is the corresponding nuclear-spin operator with $I=5/2$; $\mathcal{H}_Q$ is the nuclear quadrupole Hamiltonian for the $^{17}$O nucleus; and $\mathbf{A}^{(\mathrm{O})}$ is the hyperfine tensor.

For the c-axis-oriented PL6 (Fig.~\ref{Figure 5}(a)) and PL8$'$ (Fig.~\ref{Figure 5}(b)) centers, magnetic fields of 35 G and 105 G, respectively, were applied along the c-axis. Under these conditions, six hyperfine peaks are clearly resolved in both the left and right ODMR branches, in good agreement with the expected splitting pattern for a coupled nuclear spin with $I=5/2$. The spectra can therefore be fitted using six Lorentzian components to extract the corresponding hyperfine parameters. The asymmetry in the spectral features may be related to the nuclear spin polarization. The extracted average $A_{zz}$ value is approximately 3.4~MHz for PL6 and 3.2~MHz for PL8$'$. These values are of the same order of magnitude as the theoretical predictions for the O$_\text{C}$V$_\text{Si}$ family~\cite{kobayashi2023oxygen}. We note that the quadrupole moment of $^{17}$O does not contribute to the splitting in the respective ODMR spectrum, so we applied \textit{ab initio} calculations to predict their values in the respective defect configurations (see Supplementary Information O for details). The computed quadrupole interaction constants are around 7~MHz for each configuration. Experimental identification of the strength of this interaction is beyond the scope of the present study.

In contrast, for the basal-type PL5 and PL7$'$ centers under an external magnetic field applied along the c-axis, the ODMR spectra exhibit much richer and more crowded resonance features than a simple axial six-line pattern. For basal-type defects, the c-axis magnetic field is off-axis with respect to the defect's symmetry axis, the electron--nuclear spin states become mixed, allowing otherwise weak transitions to become observable. In addition, the lower symmetry of the basal configuration generally enhances the anisotropic contributions of both the hyperfine and quadrupole interactions, further increasing the spectral complexity. Figures~\ref{Figure 5}(c) and (d) show the $^{17}$O hyperfine spectra of the left branch of the basal-type PL5 center (PL5-L) and the right branch of the basal-type PL7$'$ center (PL7$'$-R), respectively, as the external magnetic field applied along the crystal c-axis is increased from 0 to 35 G and 70 G. With increasing magnetic field, the ODMR spectra become progressively broadened and multiple additional resonance features emerge, clearly indicating the presence of hyperfine interactions.

The direct observation of $^{17}$O hyperfine coupling and the optical and spin signatures presented above---including the ZPL and ODMR characteristics of these centers---provide strong and self-consistent evidence that the observed oxygen-related modified divacancies are indeed oxygen-vacancy complexes, namely OV centers. In addition, the hyperfine couplings of PL6 to nearby $^{29}$Si and $^{13}$C nuclear spins, presented in Supplementary Information P, are in excellent agreement with the theoretical O$_\text{C}$V$_\text{Si}$ ($hh$) model~\cite{zhao2025analyzing}, further corroborating this structural assignment. This identification establishes a solid basis for future theoretical calculations, numerical modeling, defect engineering, and coherence optimization for this family of color centers.

\section{Conclusion}
In summary, we have demonstrated that oxygen-ion implantation offers an effective and controllable approach to generating {oxygen-related} modified divacancy color centers in 4H-SiC. Four distinct spin-active centers—PL5, PL6, PL7$'$, and PL8$'$—were successfully created and systematically characterized in terms of their optical and spin properties, including zero-phonon lines (ZPLs), zero-field splittings (ZFSs), magnetic-field dependence, and temperature-dependent behavior. Our results suggest that PL7$'$ likely corresponds to the previously reported PL7, and we have identified PL8$'$ as a new single-spin defect. The measured ZPLs, ZFSs, and spin coherence times agree well with theoretical predictions for oxygen-related complexes~\cite{kobayashi2023oxygen,iwamoto2024oxygen,bai2025origin}, supporting their assignment as oxygen-related defects.

{More importantly, the oxygen-vacancy structure of these defects is directly established through isotope-resolved hyperfine interactions with the $^{17}$O nuclear spin. In samples implanted with $^{17}$O ions, we successfully resolved the corresponding hyperfine splittings. Together with the excellent agreement between experiment and theory, these results provide direct microscopic evidence that the four oxygen-related modified divacancies studied here correspond to the four crystallographic configurations of O$_\text{C}$V$_\text{Si}$ centers.}

The controllable generation of {single O$_\text{C}$V$_\text{Si}$ centers} in SiC holds significant potential for single-spin quantum application. In this work, shallow O$_\text{C}$V$_\text{Si}$($hh$)-PL6 centers were successfully applied to magnetic-field sensing~\cite{li2025non}. In addition, the realization of high-concentration, high-quality ensembles, together with their demonstrated coherent spin manipulation, represents an important step toward practical solid-state quantum technologies. The superior optical and spin properties of these defects make them promising candidates for room-temperature quantum metrology and scalable quantum networks.

% Experimental section

\section{Experimental Section}
Sample preparation:
An $n$-type single-crystal 4H-SiC epitaxial layer with a thickness of approximately 11~$\mu$m was grown on a 4$^{\circ}$ off-axis 4H-SiC substrate. The nitrogen doping concentration was below $1\times10^{14}$~cm$^{-3}$. 

For single-defect preparation, oxygen ions were implanted at a dose of $1\times10^{8}$~cm$^{-2}$, followed by annealing in an {Ar} atmosphere at 1050~$^{\circ}$C. 
For ensemble-defect preparation, samples were implanted with doses ranging from $1\times10^{9}$ to $1\times10^{15}$~cm$^{-2}$ and subsequently annealed under identical vacuum conditions at temperatures of 300, 400, 500, 600, 750, 900, 1050, and 1150~$^{\circ}$C for 0.5~hours each.
{All samples were subjected to an implantation energy of 30 keV. Detailed results comparing annealing under argon and vacuum atmospheres are presented in Supplementary Information Q.}

Experimental setups:
Optical measurements were performed using a home-built confocal microscopy system. 
A 914~nm laser served as the excitation source, coupled into the setup through a 950~nm short-pass filter and a 980~nm dichroic mirror, and focused onto the sample using an objective lens with a numerical aperture of 0.85. 
Photoluminescence was collected through a 1000~nm long-pass filter and coupled into a 1060~nm single-mode fiber. 

For low-temperature measurements, the sample was mounted in a Montana Instruments cryostat equipp-\\ed with an Autocube positioning stage. 
Room-temperature imaging was performed using a piezoelectric scanner, whereas low-temperature imaging employed a two-axis galvo scanning system with silver-coated mirrors (Thorlabs, GVS012). 
Photon detection was carried out using a superconducting nanowire single-photon detector (Photon Technology Co.). 

Laser and microwave pulse sequences were modulated by an acousto-optic modulator (AOM; Gooch~\&~H-\\ousego) and microwave switches (Mini-Circuits, ZASWA-2–50DRA+), both synchronized by a pulse generator (SpinCore, PBESR-PRO-500-PCI). 
Microwave (MW) signals were generated by a signal generator (Mini-Circuits, SSG-6000-RC), amplified by a power amplifier (Mini-Circuits, ZHL-25W-272+), and delivered to the sample through a 20~$\mu$m-diameter copper wire. 

A Time Tagger device was employed for second-order correlation function ($g^{(2)}(\tau)$) measurements. 
Spectra were recorded using a custom-built single-photon spectrometer consisting of a Newport high-precision rotation stage and a diffraction grating (600~lines/mm, blaze wavelength 1000~nm, blaze angle 17.46$^{\circ}$).

\medskip
\textbf{Supporting Information} \par %Please delete the Suppporting Information statement if it is not applicable. Please supply Supporting Information in another file. Supporting information should not be provided in .tex format
Supporting Information is available from the Wiley Online Library or from the author.

\medskip
\textbf{Author Contributions Statement} \par
J.-S. X. and C.-F. L. conceived the experiments. Q.-C. H. built the experimental setup and carried out the measurements with assistance from J.-Y. Z., S.-R., Z.-X. H., Z.-H. H., R.-J. L., and W.-X. L. X. H. and A. G. provided the theoretical analysis. A. G. suggested the comparison using oxygen-isotope–implanted samples. {X. H. carried out density functional theory calculations under supervision from A. G.} Q.-C. H. and J.-S. X. analyzed the data and wrote the manuscript, with input from all co-authors. J.-S. X., C.-F. L., and G.-C. G. supervised the project. All authors discussed the results.\par
% Acknowledgements
\medskip
\textbf{Acknowledgements} \par %delete if not applicable))
This work was supported by National Natural Science Foundation of China (No. W2411001 and No.\ 92365205), the Quantum Science and Technology-National Science and Technology Major Project (No.\ 2021ZD0301400 and No.\ 2021ZD0301200), the CAS Project for Young Scientists in Basic Research (YSBR-131) and USTC Major Frontier Research Program (No.\ LS2030000002). This work was partially performed at the University of Science and Technology of China Center for Micro and Nanoscale Research and Fabrication. {A. G. acknowledges the support from European Commission for the SPINUS (Grant No.\ 101135699) and QuSPARC (Grant No.\ 101186889) projects.}\par
\medskip
\textbf{Competing Interests Statement}\par
	The authors declare no competing interests.
% References
\medskip

\medskip
\textbf{Data Availability Statement}\par
	The data that support the findings of this study are available from the corresponding author upon reasonable request.
% References
\medskip

% Use the following code if you wish to generate your bibliography with BibTeX;
% replace the string "MSP-template" below with the name(s) of
% the BibTeX data base(s) you want to use.
% The resulting bibliography-output (the content of the .bbl file)
% must be pasted back into this file before submission.
% Please also include your BibTeX data base file(s) in your submission
% so that we can re-run BibTeX if necessary.
%
%\bibliographystyle{MSP}
%\bibliography{MSP-template}

% Figures/tables and captions
% Permission statements are required for all figures reproduced or adapted from previously published articles/sources. Please also ensure that all necessary permissions to reproduce images have been received
% Please remove these statements for original figures

\end{document}

\title{Supplementary Information: High-yield engineering and {identification of oxygen-related} modified divacancies in 4H-SiC}

\author{Qi-Cheng Hu}
\thanks{These authors contribute equally to this work}
\affiliation{Laboratory of Quantum Information, University of Science and Technology of China, Hefei, Anhui 230026, China}
\affiliation{Anhui Province Key Laboratory of Quantum Network, University of Science and Technology of China, Hefei, Anhui 230026, China}
\affiliation{CAS Center For Excellence in Quantum Information and Quantum Physics, University of Science and Technology of China, Hefei, Anhui 230026, China}
\affiliation{Hefei National Laboratory, University of Science and Technology of China, Hefei, Anhui 230088, China}

\author{Ji-Yang Zhou}
\thanks{These authors contribute equally to this work}
\affiliation{Laboratory of Quantum Information, University of Science and Technology of China, Hefei, Anhui 230026, China}
\affiliation{Anhui Province Key Laboratory of Quantum Network, University of Science and Technology of China, Hefei, Anhui 230026, China}
\affiliation{CAS Center For Excellence in Quantum Information and Quantum Physics, University of Science and Technology of China, Hefei, Anhui 230026, China}

\author{Shuo Ren}
\affiliation{Laboratory of Quantum Information, University of Science and Technology of China, Hefei, Anhui 230026, China}
\affiliation{Anhui Province Key Laboratory of Quantum Network, University of Science and Technology of China, Hefei, Anhui 230026, China}
\affiliation{CAS Center For Excellence in Quantum Information and Quantum Physics, University of Science and Technology of China, Hefei, Anhui 230026, China}

\author{Zhen-Xuan He}
\affiliation{Laboratory of Quantum Information, University of Science and Technology of China, Hefei, Anhui 230026, China}
\affiliation{Anhui Province Key Laboratory of Quantum Network, University of Science and Technology of China, Hefei, Anhui 230026, China}
\affiliation{CAS Center For Excellence in Quantum Information and Quantum Physics, University of Science and Technology of China, Hefei, Anhui 230026, China}
\affiliation{Hefei National Laboratory, University of Science and Technology of China, Hefei, Anhui 230088, China}

\author{Zhi-He Hao}
\affiliation{Laboratory of Quantum Information, University of Science and Technology of China, Hefei, Anhui 230026, China}
\affiliation{Anhui Province Key Laboratory of Quantum Network, University of Science and Technology of China, Hefei, Anhui 230026, China}
\affiliation{CAS Center For Excellence in Quantum Information and Quantum Physics,
University of Science and Technology of China, Hefei, Anhui 230026, China}
   
\author{Rui-Jian Liang}
\affiliation{Laboratory of Quantum Information, University of Science and Technology of China, Hefei, Anhui 230026, China}
\affiliation{Anhui Province Key Laboratory of Quantum Network, University of Science and Technology of China, Hefei, Anhui 230026, China}
\affiliation{CAS Center For Excellence in Quantum Information and Quantum Physics, University of Science and Technology of China, Hefei, Anhui 230026, China}
	
\author{Wu-Xi Lin}
\affiliation{Laboratory of Quantum Information, University of Science and Technology of China, Hefei, Anhui 230026, China}
\affiliation{Anhui Province Key Laboratory of Quantum Network, University of Science and Technology of China, Hefei, Anhui 230026, China}
\affiliation{CAS Center For Excellence in Quantum Information and Quantum Physics, University of Science and Technology of China, Hefei, Anhui 230026, China}
\affiliation{Hefei National Laboratory, University of Science and Technology of China, Hefei, Anhui 230088, China}

\author{Xiangru Han}
\affiliation{HUN-REN Wigner Research Centre for Physics, Institute for Solid State Physics and Optics, P.O.\ Box 49, H-1525 Budapest, Hungary}

\author{Adam Gali}
\affiliation{HUN-REN Wigner Research Centre for Physics, Institute for Solid State Physics and Optics, P.O.\ Box 49, H-1525 Budapest, Hungary}
\affiliation{Department of Atomic Physics, Institute of Physics, Budapest University of Technology and Economics, M\H{u}egyetem rakpart 3., H-1111 Budapest, Hungary}
\affiliation{MTA-WFK "Lend\"ulet" Momentum Semiconductor Nanostructures Research Group, P.O.\ Box 49, H-1525 Budapest, Hungary}
   
\author{Jin-Shi Xu}
\altaffiliation{Email: jsxu@ustc.edu.cn}
\affiliation{Laboratory of Quantum Information, University of Science and Technology of China, Hefei, Anhui 230026, China}
\affiliation{Anhui Province Key Laboratory of Quantum Network, University of Science and Technology of China, Hefei, Anhui 230026, China}
\affiliation{CAS Center For Excellence in Quantum Information and Quantum Physics, University of Science and Technology of China, Hefei, Anhui 230026, China}
\affiliation{Hefei National Laboratory, University of Science and Technology of China, Hefei, Anhui 230088, China}

\author{Chuan-Feng Li}
\altaffiliation{Email: cfli@ustc.edu.cn}
\affiliation{Laboratory of Quantum Information, University of Science and Technology of China, Hefei, Anhui 230026, China}
\affiliation{Anhui Province Key Laboratory of Quantum Network, University of Science and Technology of China, Hefei, Anhui 230026, China}
\affiliation{CAS Center For Excellence in Quantum Information and Quantum Physics, University of Science and Technology of China, Hefei, Anhui 230026, China}
\affiliation{Hefei National Laboratory, University of Science and Technology of China, Hefei, Anhui 230088, China}
 
\author{Guang-Can Guo}
\affiliation{Laboratory of Quantum Information, University of Science and Technology of China, Hefei, Anhui 230026, China}
\affiliation{Anhui Province Key Laboratory of Quantum Network, University of Science and Technology of China, Hefei, Anhui 230026, China}
\affiliation{CAS Center For Excellence in Quantum Information and Quantum Physics, University of Science and Technology of China, Hefei, Anhui 230026, China}
\affiliation{Hefei National Laboratory, University of Science and Technology of China, Hefei, Anhui 230088, China}

\maketitle
\date{\today}
\tableofcontents

\newpage

\subsection{\textcolor{black}{Comparison of samples implanted with carbon, nitrogen, and oxygen ions}}

\begin{figure*}[htbp]
\centering
\includegraphics[scale = 0.5]{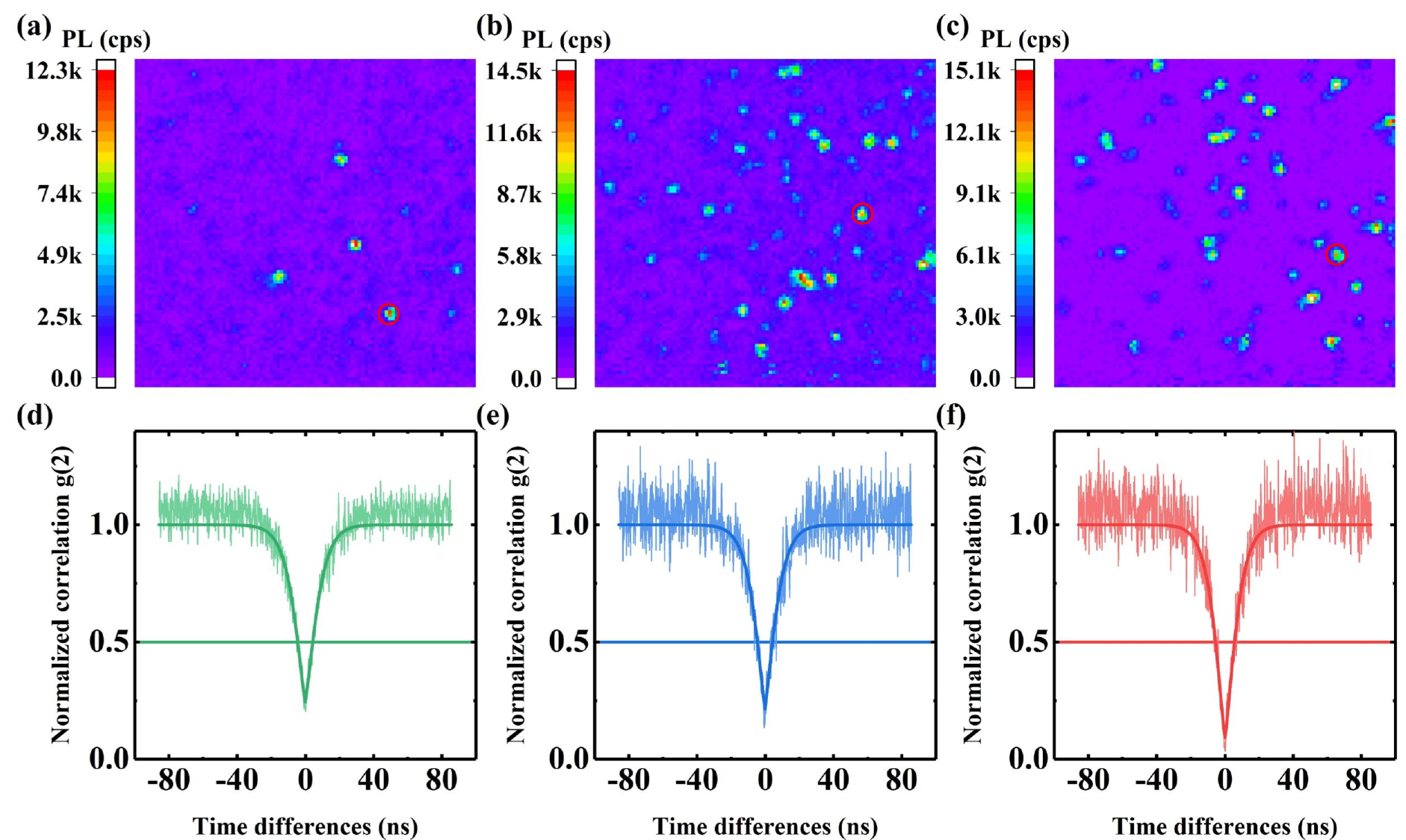}
\caption{Fluorescence scanning maps of samples implanted with different ions and the corresponding second-order correlation functions {$g^{(2)}(\tau)$} of single defects. 
(a)–(c) Scanning maps of samples implanted with carbon, nitrogen, and oxygen ions at doses of 1×10$^{9}$ cm$^{-2}$, 1×10$^{9}$ cm$^{-2}$, and 1×10$^{8}$ cm$^{-2}$, respectively. The scanning area is 20 × 20 $\mu$m$^2$, and the laser power is 200 $\mu$W. A single PL6 defect is marked with a red circle in each sample. (d)–(f) Raw second-order correlation function measurements {$g^{(2)}(\tau)$} acquired with the same laser power (200 $\mu$W) for the corresponding single defects.}
\label{S1}
\end{figure*}

Fluorescence scans were conducted on 4H-SiC samples implanted with carbon, nitrogen, and oxygen ions under a laser power of 200~$\mu$W, as shown in Fig.~S\ref{S1}. All samples were annealed in an argon (Ar) atmosphere at 1050$^\circ$C. 
Notably, the sample implanted with oxygen ions at a dose of $1\times10^{8}$~cm$^{-2}$ exhibited a comparable density of bright emission spots to the nitrogen-implanted sample at $1\times10^{9}$~cm$^{-2}$, while showing a much higher density than the carbon-implanted sample at the same dose ($1\times10^{9}$~cm$^{-2}$). 
A representative single PL6 defect in each sample is highlighted with a red circle. Second-order correlation measurements on individual PL6 defects yielded raw $g^{(2)}(0)$ values of 0.20, 0.15, and 0.04 for the carbon-, nitrogen-, and oxygen-implanted samples, respectively, without background subtraction. The lower $g^{(2)}(0)$ value observed in the oxygen-implanted sample indicates an enhanced signal-to-noise ratio of the color centers. 
These results demonstrate that oxygen-ion implantation in 4H-SiC enables the formation of bright and stable single-photon emitters {associated with oxygen-related modified divacancies}, providing a promising pathway toward room-temperature quantum photonic applications.

\subsection{Power-dependent second-order correlation function}

\begin{figure*}[htbp]
\centering
\includegraphics[scale = 0.6]{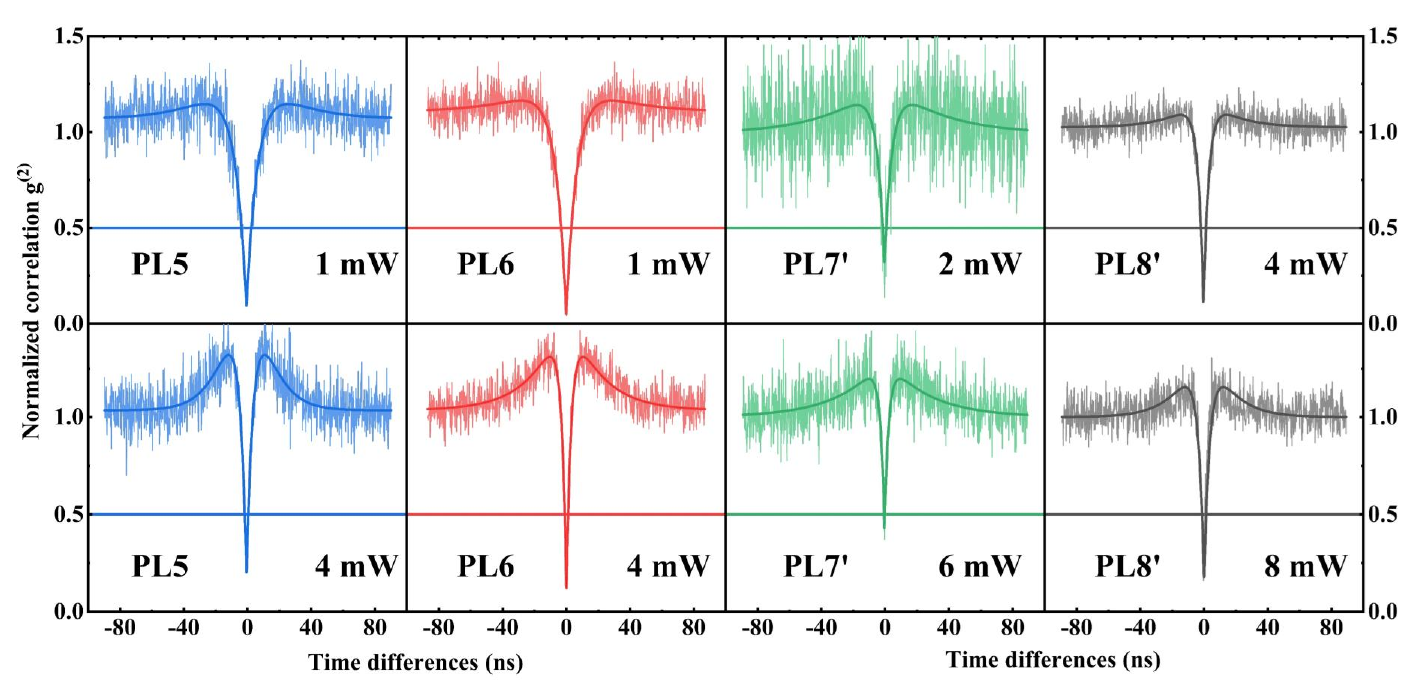}
\caption{Power-dependent second-order correlation function {$g^{(2)}(\tau)$} of single oxygen-related modified divacancies (PL5–PL8$'$) marked in Fig.~1a of the main text, measured under higher laser excitation powers.}
\label{S2}
\end{figure*}

We characterized the power-dependent second-order correlation function of single {oxygen-related} modified divacancies (PL5--PL8$'$) indicated in Fig.~1, as shown in Fig.~S\ref{S2}. As the excitation power was increased beyond 4~mW, all four types of defects exhibited a pronounced bunching effect, indicating the existence of a metastable energy level in these systems. The antibunching second-order correlation function is described by:
$
g^{(2)}(\tau) = 1 - (1 + a)e^{-|\tau - \tau_0|/t_1} + b e^{-|\tau - \tau_0|/t_2} + c
$
, where $\tau_0$ represents the time delay at zero point, and $t_1$, $t_2$, $a$, $b$, $c$ are fitting parameters. Without background subtraction, the measured $g^{(2)}(0)$ values for all these color centers remained below 0.5, confirming their single-defect nature. Among them, the PL7$'$ center exhibited a relatively lower signal-to-noise ratio due to its intrinsically weaker photoluminescence intensity. The excellent optical and spin properties of these {oxygen-related} modified divacancy centers at room temperature underscore their potential for applications in quantum sensing and quantum communication.

\subsection{Statistical analysis of room-temperature single spin defects}

\begin{figure*}[htbp]
\centering
\includegraphics[scale = 0.7]{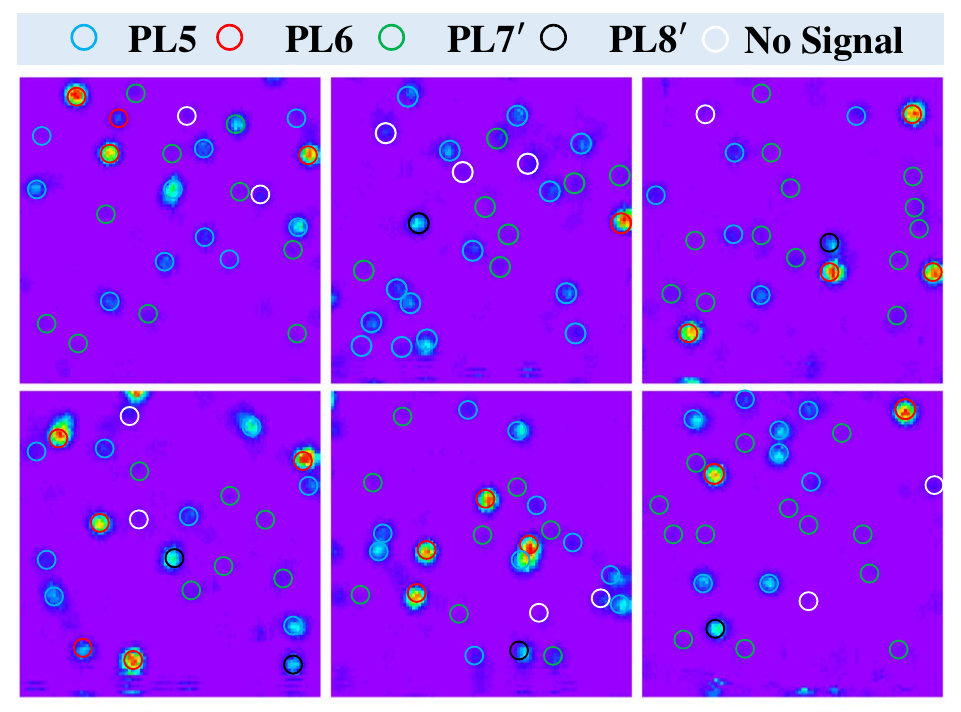}
\caption{Statistical analysis of color centers in an oxygen-ion-implanted sample. 
Six randomly selected fluorescence scan images with a scanning area of 10×10~$\mu$m$^{2}$ were analyzed. 
The blue, red, green, and black circles represent PL5, PL6, PL7$'$, and PL8$'$, respectively, while the white circles indicate single emission spots without a detectable spin signal.}
\label{S3}
\end{figure*}

We performed a statistical analysis of the {room-temperature single spin defects} generated by oxygen-ion implantation, as shown in Fig.~S\ref{S3}. 
Six fluorescence scan images were randomly selected, each covering an area of $10\times10~\mu$m$^{2}$. 
Room-temperature optically detected magnetic resonance (ODMR) measurements were carried out on all single color centers exhibiting stable photon emission in each image under zero magnetic field. 
The configurations of individual defects were identified based on their fluorescence intensity and ODMR resonant frequencies, which reflect the zero-field splitting parameters. 

In total, 149 single defects were analyzed, among which 92\% were identified as PL5, PL6, PL7$'$, or PL8$'$, while the remaining 8\% exhibited no measurable ODMR signal. 
Some of these may correspond to common divacancy centers that can only be identified at cryogenic temperatures, while others might represent single-photon emitters without spin contrast. 
The corresponding histogram is presented in the main text as Fig.~1f. 
A rough estimation indicates that, for every 100~$\mu$m$^{2}$, approximately 9 PL5, 3 PL6, 9 PL7$'$, and 1 PL8$'$ defects are formed. 
Considering an implanted ion dose of $1\times10^{8}$~cm$^{-2}$, the yield for generating {oxygen-related} spin-active modified divacancy color centers is estimated to be about 23\%, {which is, to the best of our knowledge, among the highest values reported for ion-implantation-based generation of single spin defects in SiC} {\cite{yan2020room,He2024low,li2022room}}. 
These findings demonstrate that oxygen-ion implantation in 4H-SiC provides an exceptionally high generation yield of {oxygen-related modified divacancy centers, highlighting the important role of oxygen incorporation in the formation of this defect family.}

\subsection{Comparison of $T_2$ for PL6 color centers generated by carbon, nitrogen, and oxygen ion implantation}

\begin{figure*}[htbp] 
\centering 
\includegraphics[scale=0.5]{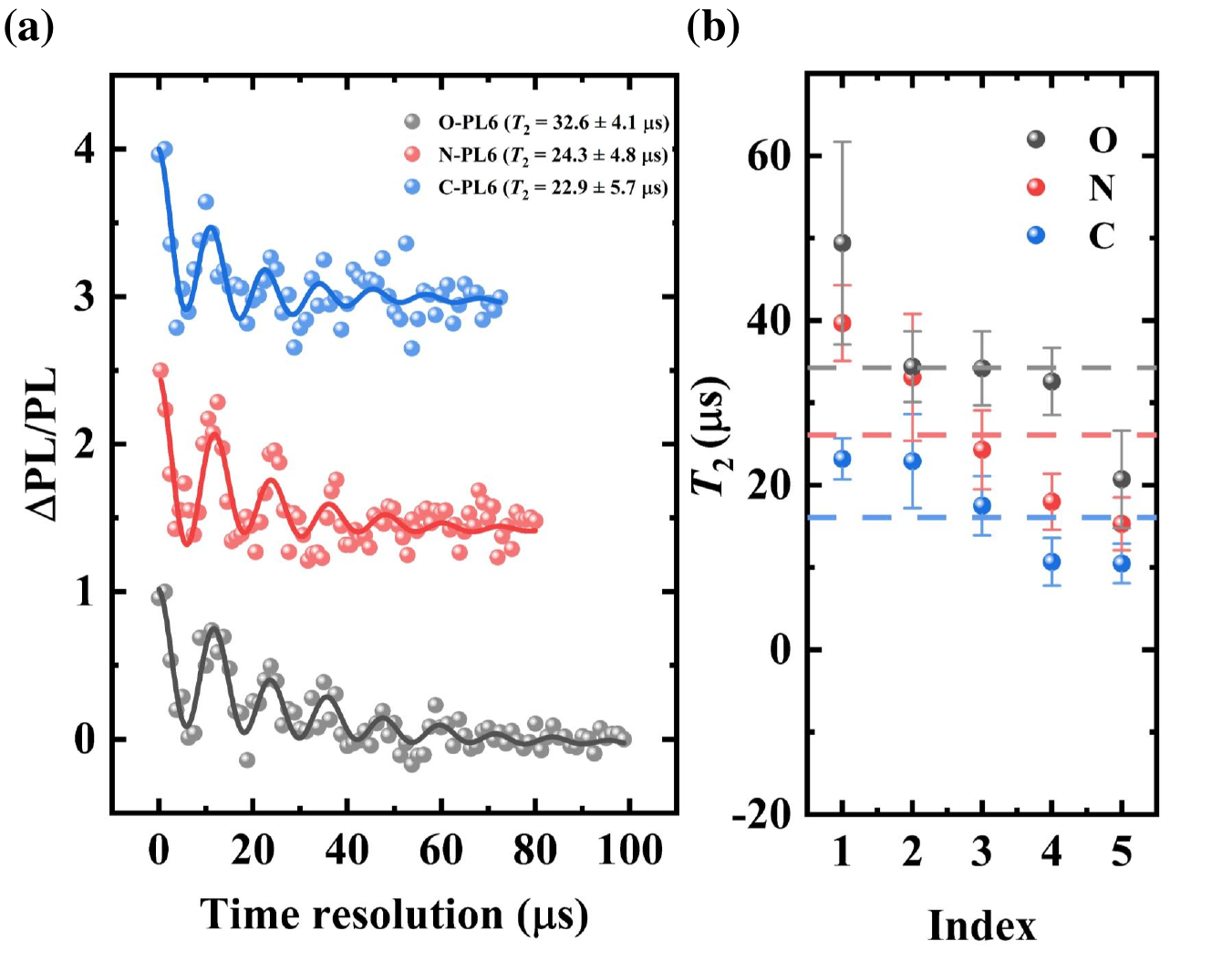} 
\caption{{Comparison of $T_2$ for PL6 color centers generated by carbon/nitrogen ($1\times10^9$ cm$^{-2}$) and oxygen ($1\times10^8$ cm$^{-2}$) ion implantation. (a) Representative $T_2$ measurements for carbon-, nitrogen-, and oxygen-ion-implanted samples obtained at 180 G. (b) Statistical analysis of $T_2$ values extracted from five randomly selected spots for each implanted species.}} 
\label{S4} 
\end{figure*}

Benefiting from the high conversion efficiency of oxygen-related modified divacancies under oxygen-ion implantation, both lattice damage caused by heavy ion bombardment and the introduction of additional impurity defects or charge traps around the color centers can be reduced. These factors, in turn, affect the relaxation and decoherence processes of the defects. Taking PL6 as an example, we performed $T_2$ measurements and a statistical analysis, as shown in Fig.~S\ref{S4}. The results indicate that the average $T_2$ coherence time of the oxygen-ion-implanted sample (34.3~$\mu$s) is enhanced by approximately 110\% and 30\% compared with those of the carbon- and nitrogen-ion-implanted sample (16.1~$\mu$s and 26.1~$\mu$s), respectively. The relatively modest improvement is likely limited by the abundant $^{13}$C and $^{29}$Si nuclear spins in SiC, whose spin bath dominates the decoherence process. Future work employing dynamic decoupling techniques may further extend the coherence time.

\subsection{Spin coherent control of individual PL7$'$ and PL8$'$ centers}

\begin{figure*}[htbp]
\centering
\includegraphics[scale = 0.5]{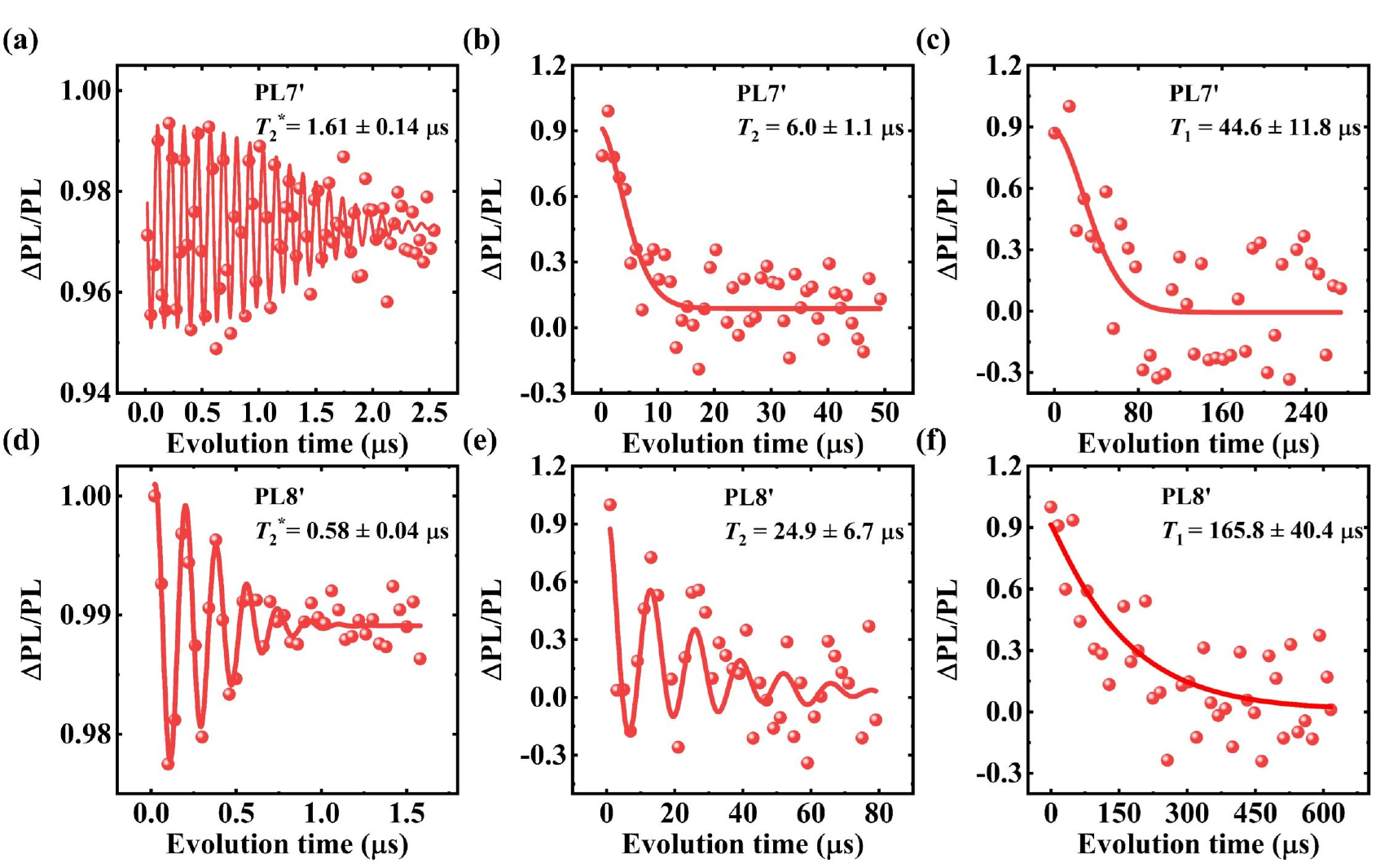}
\caption{Spin coherence control of individual PL7$'$ and PL8$'$. (a) and (d) Ramsey inhomogeneous dephasing time ($T_2^*$) measurements of PL7$'$ without an external magnetic field, and of PL8$'$ at 180~G. (b) and (e) Hahn-echo spin coherence time ($T_2$) measurements of PL7$'$ (left-branch resonance) without an external magnetic field and PL8$'$ (left-branch resonance) under a magnetic field of 180~G. (c) and (f) Spin-lattice relaxation time ($T_1$) measurements were performed on PL7$'$ without an external magnetic field and on PL8$'$ at 180~G.}
\label{S5}
\end{figure*}

We further realized spin-coherence control of individual PL7$'$ and PL8$'$ centers by measuring their Ramsey inhomogeneous dephasing time ($T_2^*$), Hahn-echo coherence time ($T_2$), and spin-lattice relaxation time ($T_1$). The basal-oriented PL7$'$ color center hosts {mixed} $m_{s}=\pm 1$ energy levels, enabling coherent manipulation from $\ket{0}$ to $\ket{+}=1/\sqrt{2}(\ket{1}+\ket{-1})$ via microwave excitation, similar to PL5. The c-axis-oriented PL8$'$ color center exhibits degeneracy lifting of the $m_{s}=\pm 1$ levels under an applied magnetic field along the c-axis. Accordingly, we also performed coherent measurements on the $\ket{0}$ to $\ket{1}$ transition under a magnetic field of 180~G. 

The experimental results are shown in Fig.~S\ref{S5}. For PL7$'$ (Fig.~S\ref{S5}a--c), the measured values are $T_2^* = 1.61 \pm0.14~\mu$s, $T_2 = 6.0\pm1.1~\mu$s, and $T_1 = 44.6 \pm 11.8~\mu$s. The $T_2$ coherence time of the left branch of PL7$'$ (6.0 $\pm~1.1$ $\mu$s) is shorter than the value previously reported for the right branch~\cite{li2022room}. For PL8$'$ (Fig.~S\ref{S5}d--f), the corresponding values are $T_2^* = 0.58 \pm 0.04~\mu$s, $T_2 = 24.9 \pm 6.7~\mu$s, and $T_1 = 165.8 \pm 40.4~\mu$s.

\subsection{SRIM simulation of vacancy density and depth distribution}

\begin{figure*}[htbp]
\centering
\includegraphics[scale = 0.5]{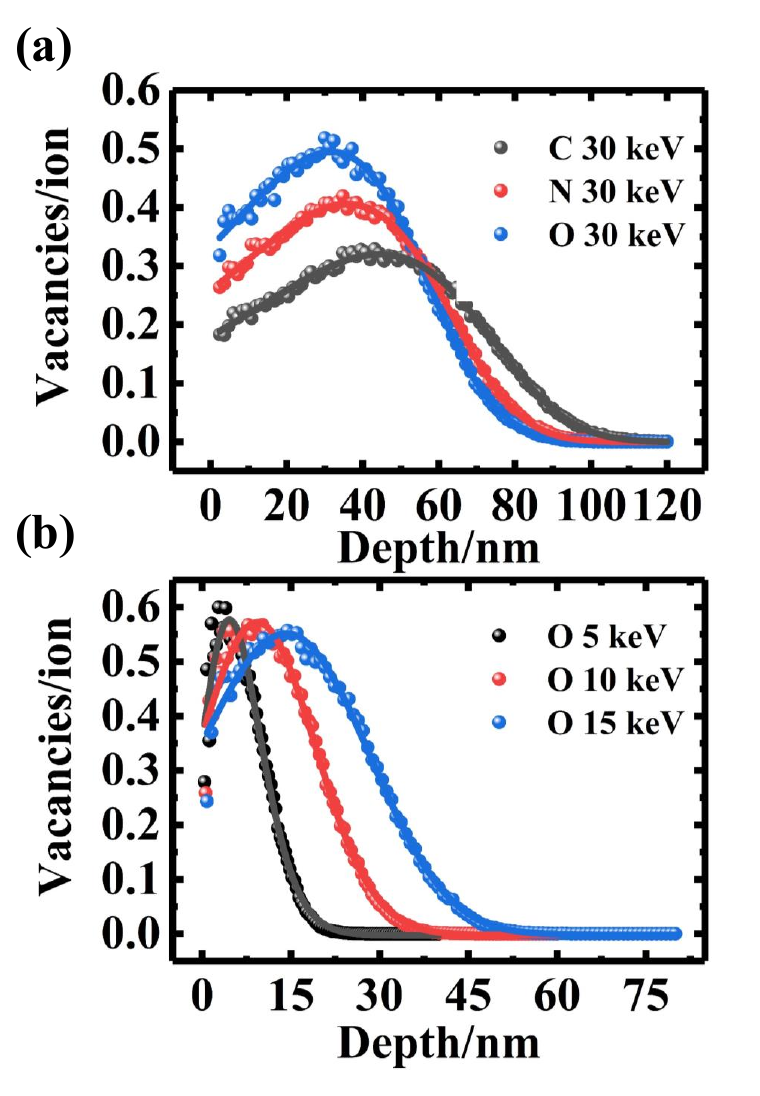}
\caption{SRIM simulation of ion implantation in 4H-SiC. 
(a) Vacancy density and depth distribution for carbon, nitrogen, and oxygen ion implantation at 30~keV. 
(b) Vacancy density and depth distribution for oxygen ion implantation at 5, 10, and 15~keV.}
\label{S6}
\end{figure*}

We performed Stopping and Range of Ions in Matter (SRIM) simulations for carbon-, nitrogen-, and oxygen-ion implantation in 4H-SiC. 
The simulated vacancy concentration and depth profiles at an implantation energy of 30~keV are shown in Fig.~S\ref{S6}a. 
The maximum vacancy depths for carbon, nitrogen, and oxygen ion implantation are 44.5~nm, 35.9~nm, and 31.8~nm, respectively. 
Owing to their heavier mass, oxygen ions exhibit faster energy loss and consequently shallower penetration depths. 
The average number of vacancies produced per implanted ion was found to be 216, 240, and 270 for carbon, nitrogen, and oxygen ions, respectively, indicating that the total number of vacancies generated by different ions is {of the same order}. Previous theoretical studies have shown that the formation energy of oxygen--vacancy complexes is lower than that of pristine divacancies~\cite{bai2025origin,kobayashi2023oxygen}. 
Combined with the experimentally observed high generation yield in oxygen-ion implantation, this suggests that oxygen plays a crucial role in the formation of {oxygen-related} modified divacancy centers. 

We further simulated oxygen-ion implantation at different energies, as shown in Fig.~S\ref{S6}b. 
While the maximum vacancy concentration remains nearly constant, the depth distribution of vacancies strongly depends on the implantation energy. 
The mean implantation depths were 4.6~nm, 9.3~nm, and 14.2~nm for oxygen-ion energies of 5, 10, and 15~keV, respectively. 
The results indicate that higher implantation energies lead to deeper penetration depths but also cause broader vacancy distribution profiles. 
These findings provide valuable guidance for engineering color centers at controlled depths to optimize coupling with micro- and nano-structured optical modes.

\subsection{Demonstration of sensing with shallow PL6 centers}
\begin{figure}[htbp]
\centering
\includegraphics[scale = 0.5]{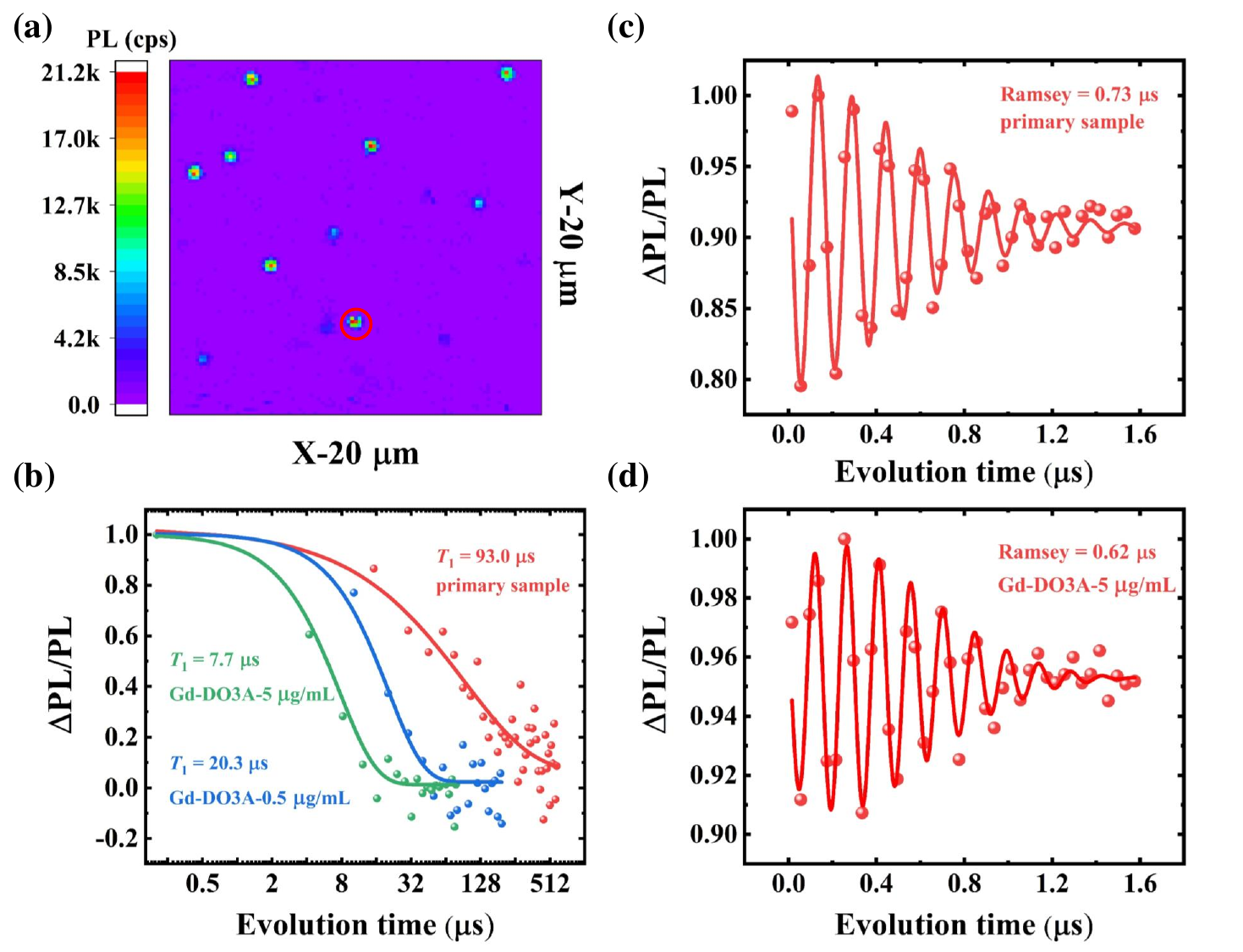}
\caption{
Demonstration of sensing with shallow PL6 prepared by oxygen-ion implantation. 
(a) Confocal photoluminescence (PL) scan of a sample implanted with oxygen ions (5~keV, dose: $1\times10^8$~ions/cm$^2$). A single PL6 emitter, marked by a red circle, was randomly selected and tracked throughout subsequent surface processing. 
(b) Measured spin-lattice relaxation times ($T_1$) for the primary sample and after treatment with Gd-DO3A solutions of different concentrations. The horizontal axis is plotted on a base-2 logarithmic scale. 
(c, d) Ramsey measurement results for the primary sample (c) and after treatment with a high-concentration Gd-DO3A solution (d).}
\label{S7}
\end{figure}

Shallow oxygen-related modified divacancy centers, owing to their excellent charge stability and controllable spin states at room temperature, hold great promise for near-surface quantum sensing applications.
Single {oxygen-related} modified divacancies were created by implanting 5~keV oxygen ions at a dose of $1\times10^{8}$~cm$^{-2}$.
According to SRIM simulations, the average depth of the generated defects is approximately 4.6~nm.
Under 914~nm laser excitation with a power of 0.2~mW, confocal PL imaging revealed stable single-photon emission from individual PL6 centers.
A representative PL6 single-photon emitter is highlighted in Fig.~S\ref{S7}a, where the spin-lattice relaxation time ($T_1$) was characterized before and after surface treatment with Gd-DO3A solutions.
As shown in Fig.~S\ref{S7}b, the $T_1$ time decreased markedly from 93~$\mu$s to 7.7~$\mu$s as the Gd-DO3A concentration increased from 0.5 to 5~$\mu$g/mL, indicating enhanced magnetic noise induced by surface paramagnetic ions.
Furthermore, Ramsey measurements (Fig.~S\ref{S7}c,d) revealed a reduction in the inhomogeneous dephasing time ($T_2^*$) from 0.73~$\mu$s to 0.62~$\mu$s after surface treatment. These results agree well with our previous {work~\cite{li2025non}}.
\\
\\
\\
\\
\\
\subsection{The basic parameters of different centers and schematic diagrams of the oxygen-vacancy centers.}
\begin{figure}[htbp]
\centering
\includegraphics[scale = 0.7]{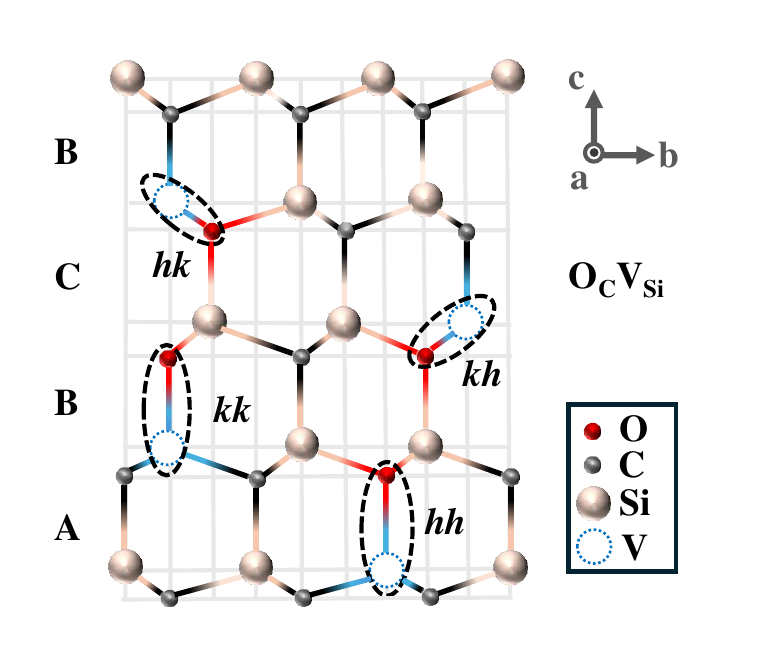}
\caption{Schematic diagram of the oxygen-vacancy structure in 4H-SiC.}
\label{S16}
\end{figure}

Based on previous studies and the experimental results presented in this work, we summarize the basic parameters of divacancies and modified divacancies in 4H-SiC in Supplementary Table 1, including the atomic lattice, orientation, zero-phonon line (ZPL), zero-field splitting (ZFS) at room temperature (RT), and ZFS at low temperature (LT). In addition, schematic diagrams of the oxygen-vacancy structures are presented in Fig.~S\ref{S16}.
\begin{table*}[!htbp]
\centering
\caption{\fontsize{10pt}{16pt}\selectfont {Summary of PL1--PL8 color centers in 4H-SiC}}
\begin{threeparttable}
\scalebox{1.3}{
\begin{tabular}{cccccccc}
\hline
Types & Lattice & Orientation & ZPL (nm) & \multicolumn{2}{c}{ZFS at RT (MHz)} & \multicolumn{2}{c}{ZFS at LT (MHz)}\\
\hline
PL1~\cite{koehl2011room,falk2013polytype,li2022room} & V$_\text{C}$V$_\text{Si}$($hh$) & c-axis & 1132 & \multicolumn{2}{c}{1323.5} & \multicolumn{2}{c}{1336}\\
PL2~\cite{koehl2011room,falk2013polytype,li2022room} & V$_\text{C}$V$_\text{Si}$($kk$) & c-axis & 1131 & \multicolumn{2}{c}{NO} & \multicolumn{2}{c}{1305}\\
PL3~\cite{koehl2011room,falk2013polytype,li2022room} & V$_\text{C}$V$_\text{Si}$($hk$) & basal & 1108 & 1135 & NO & 1141 & 1305\\
PL4~\cite{koehl2011room,falk2013polytype,li2022room} & V$_\text{C}$V$_\text{Si}$($kh$) & basal & 1078 & NO & 1334 & 1317 & 1354\\
PL5~\cite{koehl2011room,falk2013polytype,li2022room} & {O$_\text{C}$V$_\text{Si}$($hk$)\tnote{2}} & basal & 1042 & 1343 & 1375 & 1357 & 1388\\
PL6~\cite{koehl2011room,falk2013polytype,li2022room} & {O$_\text{C}$V$_\text{Si}$($hh$)\tnote{2}} & c-axis & 1038 & \multicolumn{2}{c}{1351} & \multicolumn{2}{c}{1365}\\
PL7~\cite{falk2013polytype,li2022room}/PL7$'$& {O$_\text{C}$V$_\text{Si}$($kh$)\tnote{2}} & basal & 1106\tnote{1} & 1135\tnote{1} & 1333 & 1141(40~K)\tnote{1} & 1348(40~K)\tnote{1}\\
PL8$'$ & {O$_\text{C}$V$_\text{Si}$($kk$)\tnote{2}} & c-axis & 1077\tnote{1} & \multicolumn{2}{c}{1316\tnote{1}} & \multicolumn{2}{c}{1329\tnote{1}}\\
PL8\cite{yan2020room} & Unknown & c-axis & 1007 & \multicolumn{2}{c}{1388} & \multicolumn{2}{c}{1399}\\
\hline
\end{tabular}}
\begin{tablenotes}
\footnotesize
\item[1] Identified in this work.
\item[2] The lattice directions are assigned according to the ZPL calculations reported in Ref. \cite{bai2025origin}.
\end{tablenotes}
\end{threeparttable}
\end{table*}

\subsection{Optimization of the generation of silicon vacancies and {oxygen-related} modified divacancies}
\begin{figure}[htbp]
\centering
\includegraphics[scale = 0.5]{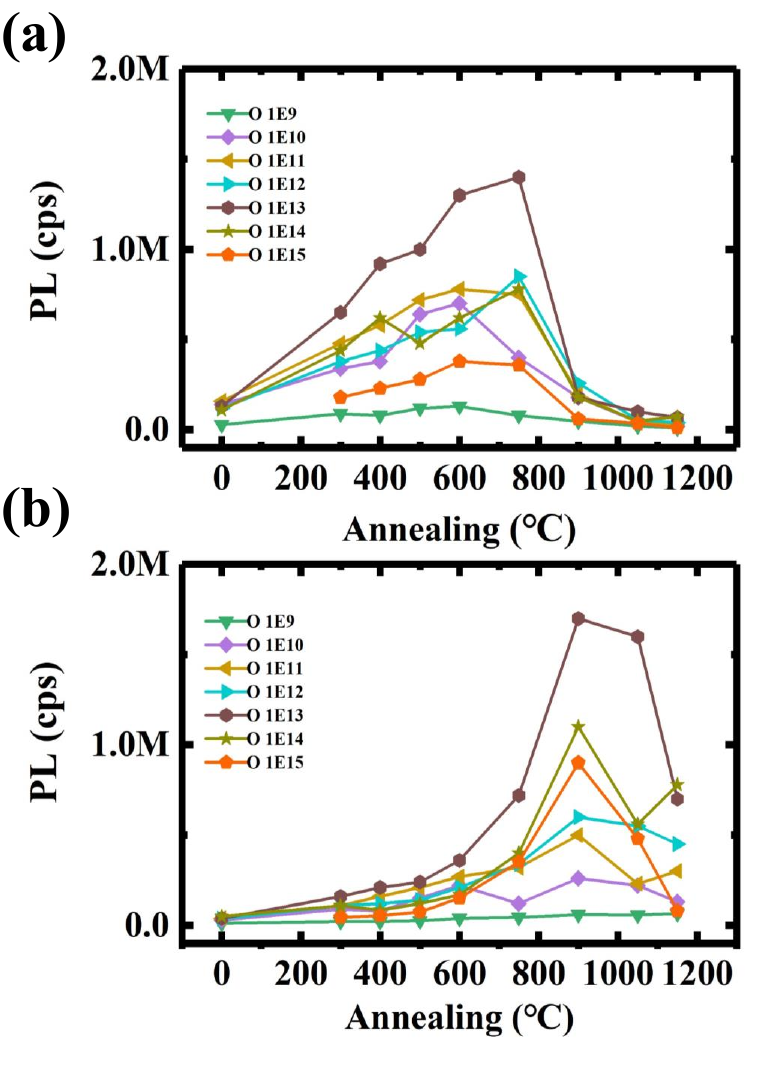}
\caption{Fluorescence counts of samples implanted with different oxygen ion doses and annealed at various temperatures. 
(a) Fluorescence counts of silicon vacancies in samples annealed from 300 to 1150~$^\circ$C and excited with a 723~nm laser. 
(b) Fluorescence counts of oxygen-related modified divacancies excited with a 914~nm laser for samples annealed from 300 to 1150~$^\circ$C.}
\label{S8}
\end{figure}

Fig.~S\ref{S8}a shows the photoluminescence (PL) intensity of silicon vacancy (V$_{\text{Si}}$) centers, measured under a laser power of 0.1~mW using an 810~nm dichroic mirror and an 850~nm long-pass filter. The PL intensity exhibits a clear dependence on the annealing temperature. For all ion implantation doses, the fluorescence signal increases with temperature up to approximately 600$^\circ$C, beyond which it gradually decreases. This behavior indicates that moderate thermal annealing effectively activates V$_{\text{Si}}$ centers by repairing implantation-induced lattice damage, whereas excessive annealing may cause vacancy diffusion or recombination, thereby reducing the concentration of optically active centers. The optimal annealing temperature for silicon vacancies is thus determined to be around 600$^\circ$C, independent of the implanted oxygen-ion dose.

The maximum PL intensity is obtained at an implantation dose of $1\times10^{13}$~cm$^{-2}$. Further increasing the implantation dose leads to a reduction in PL intensity, likely due to lattice damage caused by excessive ion bombardment, which partially amorphizes the SiC crystal. A similar phenomenon has been observed for silicon vacancies generated by carbon-ion implantation~\cite{wang2019generation}.

Fig.~S\ref{S8}b presents the PL intensity of {oxygen-related modified} divacancy centers, measured with a laser power of 0.1~mW, a 914~nm dichroic mirror, and a 1000~nm long-pass filter. The fluorescence intensity of {these centers} exhibits a distinct temperature dependence compared with V$_{\text{Si}}$ centers. The optimal annealing temperature is found to be approximately 900$^\circ$C, consistent with previous results obtained for carbon-ion implantation~\cite{li2022room,lee2021stability}.

These results demonstrate that the formation and optical activation of color centers in 4H-SiC can be effectively tuned by optimizing both the implantation dose and the annealing temperature. While silicon vacancies exhibit optimal emission at moderate annealing temperatures, the formation of {oxygen-related} modified divacancies such as PL5--PL8$'$ requires higher thermal activation, {likely to facilitate} defect migration and stabilize complex vacancy configurations.

\subsection{Influence of nitrogen doping on ion-induced {oxygen-related} modified divacancies}

\begin{figure}[htbp]
\centering
\includegraphics[scale=0.5]{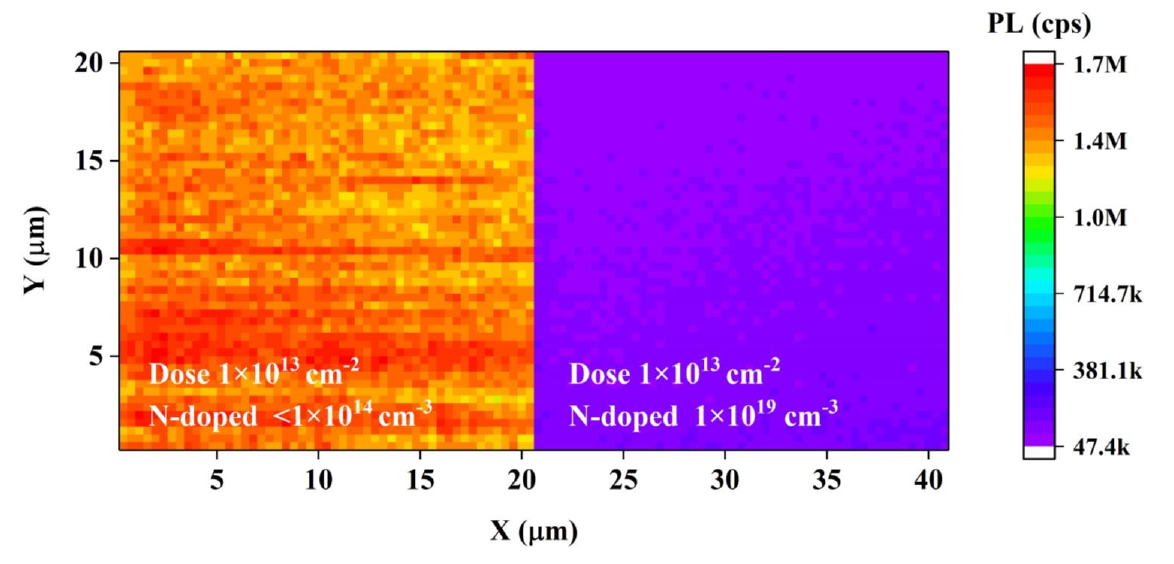}
\caption{
Effect of nitrogen doping on the fluorescence of {oxygen-related} modified divacancies. 
}
\label{S9}
\end{figure}

The nitrogen doping level in 4H-SiC strongly affects the charge-state environment surrounding color centers. 
A high nitrogen concentration typically introduces an excess of negatively charged donors, which suppresses the formation of neutral modified divacancy. To examine this effect, we compared samples with nitrogen concentrations below $1\times10^{14}$~cm$^{-3}$ and at $1\times10^{19}$~cm$^{-3}$, both subjected to oxygen-ion implantation. 
Fig.~S\ref{S9} presents confocal photoluminescence (PL) scans after implantation at doses of $1\times10^{13}$~cm$^{-2}$. 
The sample with higher nitrogen doping exhibits a markedly lower density of emissive centers compared with the lightly doped sample. 
{These results clearly indicate that the observable fluorescence from oxygen-related modified divacancies is strongly reduced in highly nitrogen-doped 4H-SiC, possibly because the relevant emissive charge state becomes less stable.}

\subsection{Demonstration of spin control in ensemble PL5 and PL6 centers}

\begin{figure*}[htbp] 
\centering \includegraphics[scale=0.7]{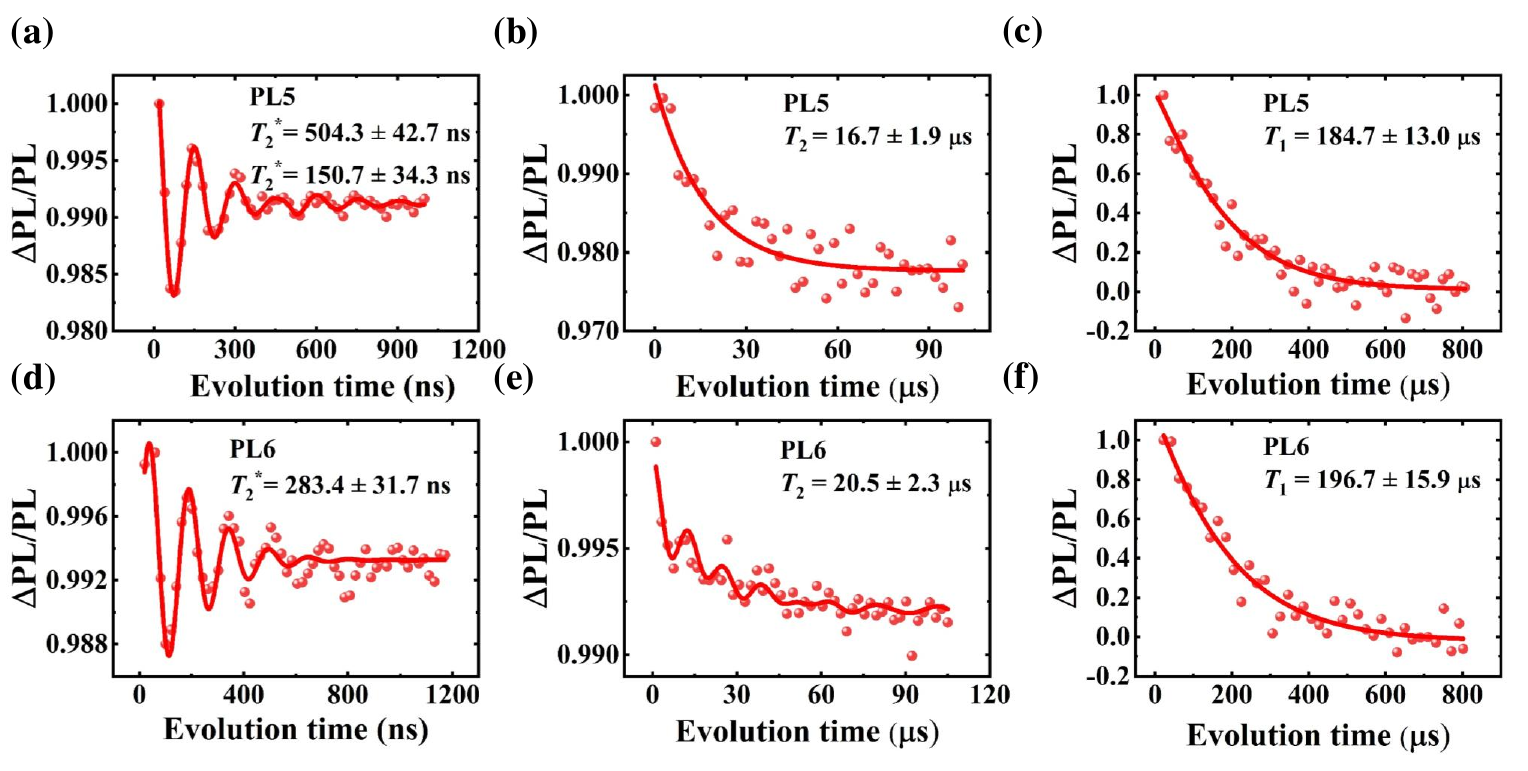} \caption{ Coherent property measurements of oxygen-ion-implanted ensemble samples (PL5 and PL6) with an implantation dose of 1×10$^{10}$ cm$^{-2}$. (a)–(c) Inhomogeneous dephasing time ($T_2^*$), spin coherence time ($T_2$), and spin–lattice relaxation time ($T_1$) measurements of PL5 without an external magnetic field. (d)–(f) Corresponding measurements for PL6 at 180 G. } \label{S10} 
\end{figure*}

We investigated the spin coherence properties of PL5 without an external magnetic field and PL6 under an external magnetic field of 180 G applied along the c-axis.
Fig.~S\ref{S10} summarizes the extracted $T_2^*$, $T_2$, and $T_1$ values.
As shown above and in Fig.~4 of the main text, the Rabi oscillation measurements of PL5 centers exhibited characteristic beating patterns.
Similarly, the Ramsey fringes reproduced comparable dual-frequency modulations, as shown in Fig.~S\ref{S10}a.
The inhomogeneous dephasing time $T_2^*$ was obtained by fitting the data with a cosine function modulated by {a double-exponential decay envelope}, yielding two characteristic components of 504.3~ns and 150.7~ns.
The ensemble PL5 sample exhibited a spin coherence time of $T_2 = 16.7~\mu$s (Fig.~S\ref{S10}b) and a spin--lattice relaxation time of $T_1 = 184.7~\mu$s (Fig.~S\ref{S10}c).

For the PL6 centers, the measured $T_2^*$ at 180~G was about 283~ns (Fig.~S\ref{S10}d), with $T_2 \approx 20.5~\mu$s (Fig.~S\ref{S10}e) and $T_1 \approx 196.7~\mu$s (Fig.~S\ref{S10}f).
{The coherence times of the ensemble centers are generally shorter than those of the corresponding single defects, primarily because of inhomogeneous broadening and dipolar spin--spin interactions, which accelerate dephasing across the ensemble.} {Additional contributions from microwave and optical field inhomogeneity, residual lattice damage, and uncorrelated local noise further reduce the observed coherence times.}
\\
\\
\\
\\
\\
\\
\\
\\
\\
\\

\subsection{Magnetic-field and Temperature-dependent behavior of ensemble defects}

\begin{figure*}[htbp]
\centering
\includegraphics[scale=0.5]{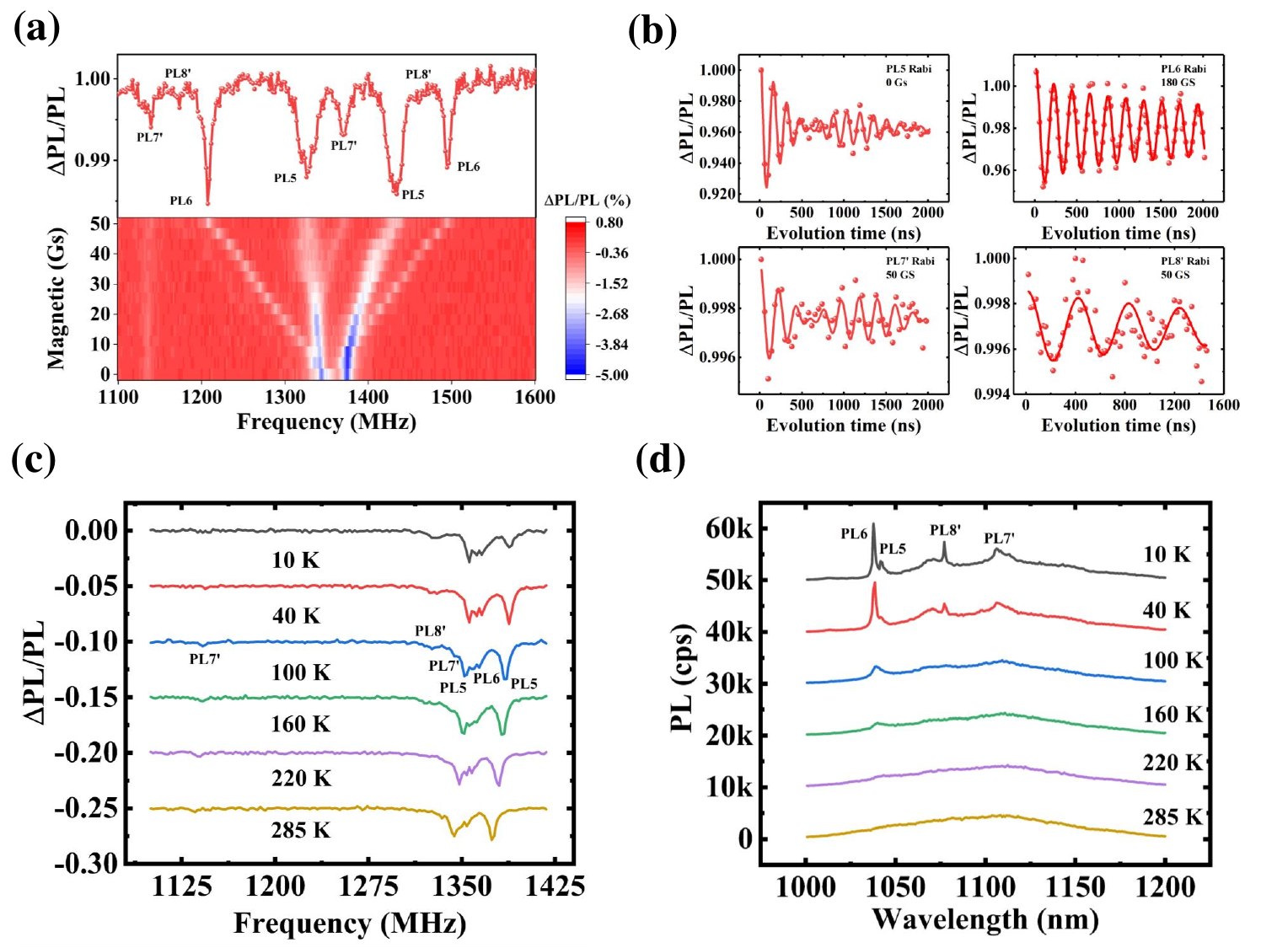}
\caption{
Spin and optical property measurements of oxygen-ion-implanted ensemble samples with an implantation dose of 1×10$^{13}$ cm$^{-2}$.
(a) Magnetic-field-dependent ODMR measurements.
(b) Rabi oscillations of PL5–PL8$'$.
(c) Temperature-dependent ODMR measurements.
(d) Temperature-dependent ZPL spectra.
}
\label{S11}
\end{figure*}

We further characterized the spin and optical properties of oxygen-ion-implanted ensemble samples. 
Fig.~S\ref{S11}a shows the magnetic-field-dependent ODMR spectra of the ensemble, where four distinct spin defects—PL5, PL6, PL7$'$, and PL8$'$—are identified, in good agreement with the single-defect results presented in the main text. The ODMR spectrum at 50~G is shown in the upper panel.

Fig.~S\ref{S11}b presents the Rabi oscillations of these spin defects. For the basal-type defects (PL5 and PL7$'$), clear beating patterns are observed, originating from the different orientations of the basal defects. In contrast, the $c$-axis--oriented defects (PL6 and PL8$'$) exhibit no beating behavior.

Fig.~S\ref{S11}c shows the temperature-dependent ODMR measurements of the ensemble sample, which display {contrast variations similar to those observed in the corresponding single-defect measurements.}
The temperature-dependent zero-phonon-line (ZPL) spectra are shown in Fig.~S\ref{S11}d, where modified divacancy defects exhibit well-resolved ZPLs below 40~K, while the ZPLs of PL1--PL2 defects are nearly invisible.

\subsection{Comparison of low-temperature ensemble spectra}

\begin{figure*}[htbp] 
\centering 
\includegraphics[scale=0.55]{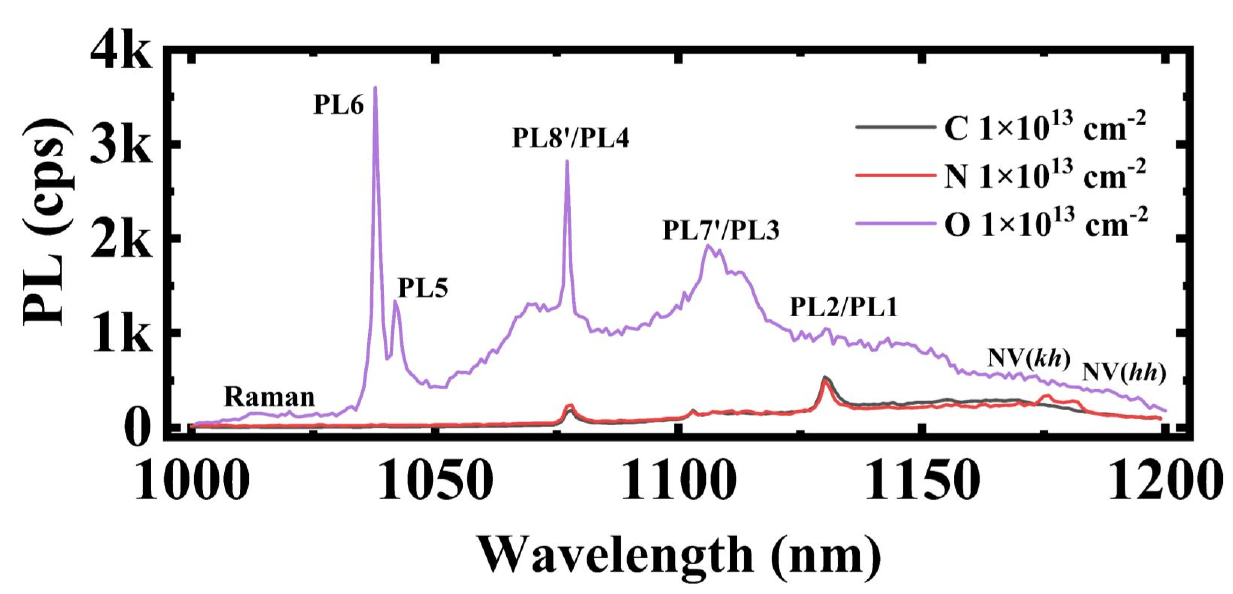} 
\caption{{Comparison of low-temperature fluorescence spectra of carbon-, nitrogen-, and oxygen-ion-implanted samples at a dose of $1\times10^{13}$ cm$^{-2}$ (8 K).}} 
\label{S12} 
\end{figure*}
According to the SRIM simulations presented in Supplementary Information F, the average numbers of vacancies generated per incident ion under 30 keV implantation are approximately 216, 240, and 270 for carbon, nitrogen, and oxygen ions, respectively. Because all implants were performed at the same fluence, the total numbers of generated vacancies are of the same order. The pronounced differences in fluorescence intensity therefore most likely originate from the formation of different defect species during annealing. As shown in Fig.~S\ref{S12}, the overall fluorescence intensity of the carbon- and nitrogen-ion-implanted samples is significantly lower than that of the oxygen-ion-implanted sample, and the relative proportion of oxygen-related modified divacancies is also much smaller. These results indicate that, under oxygen-ion implantation, the strong low-temperature fluorescence is dominated by oxygen-related modified divacancies.

\subsection{Parameters of $^{17}$O implantation}
\begin{figure*}[htbp]
\centering
\includegraphics[scale=0.5]{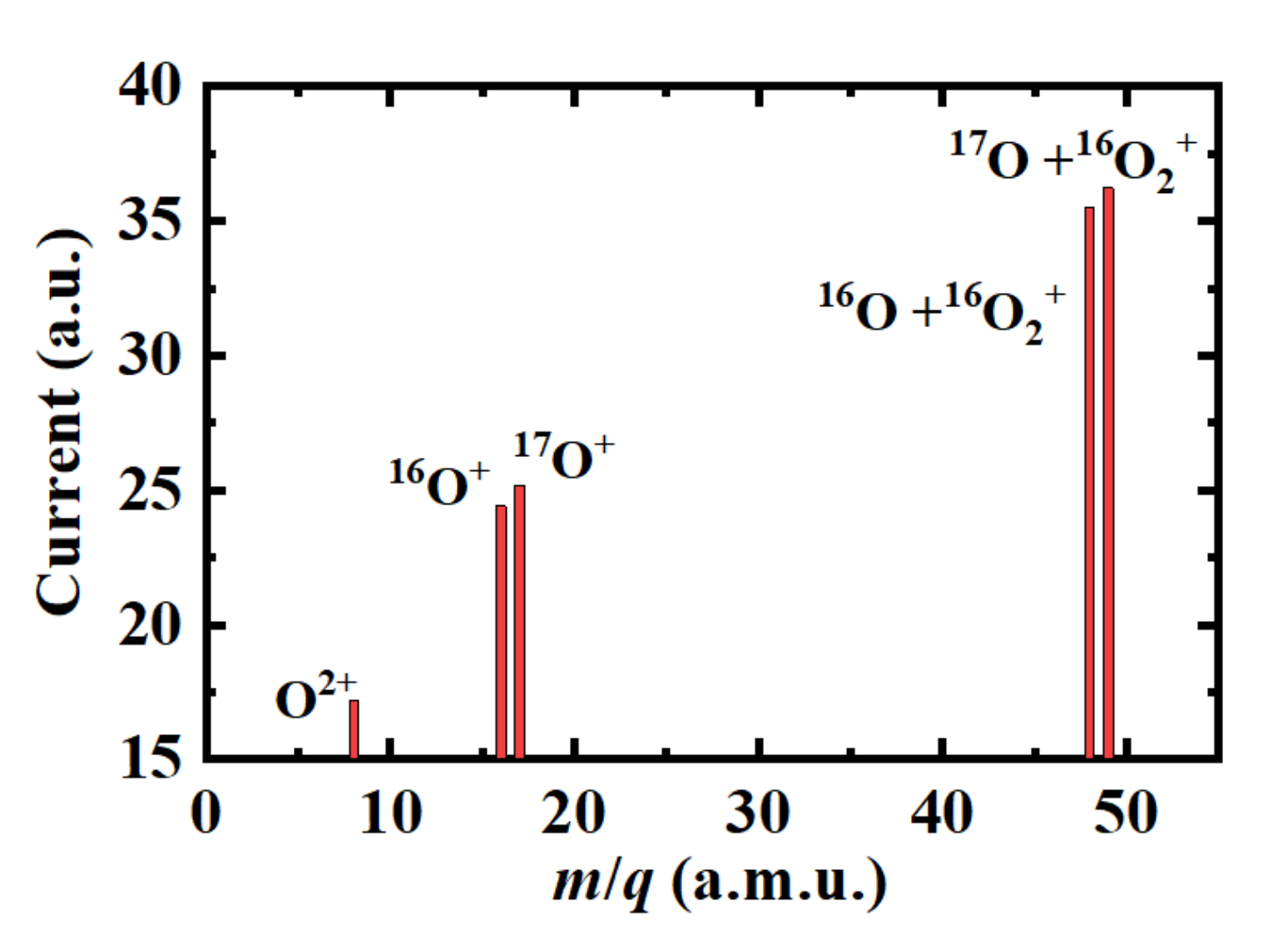}
\caption{{Ion-current spectrum for the selection of different oxygen-related ionic species during $^{17}$O implantation. Distinct peaks are resolved according to their mass-to-charge ratio ($m/q$).}}
\label{S13}
\end{figure*}

In our experiment, isotopically enriched oxygen gas with a nominal $^{17}$O abundance of 70\% was used as the implantation source. During ion implantation, different ionic species were separated in the cyclotron accelerator according to their mass-to-charge ratio ($m/q$). Fig.~S\ref{S13} shows the corresponding ion-current spectrum, in which distinct peaks can be assigned to different oxygen-related ionic species. Here, the vertical axis represents the relative ion-beam current measured during the species-selection process, and the different peak heights reflect the relative beam intensities of the corresponding ionic species in the accelerator beamline. This mass selection enables the implantation of the desired $^{17}$O-containing ions and is therefore essential for the subsequent identification of intrinsic $^{17}$O hyperfine signals.

\subsection{Quadrupole moment of $^{17}$O from first principles: methods and results}
All calculations are performed within density functional theory (DFT) using the Vienna \textit{Ab initio} Simulation Package (VASP)~\cite{kresse_efficient_1996} with the projector augmented wave (PAW) method~\cite{blochl_projector_1994}. The oxygen-vacancy defect (O$_\text{C}$V$_\text{Si}$ or simply OV) structures (Fig.~S\ref{S16}) are modeled in a 576-atom 4H-SiC supercell using $\Gamma$-point sampling. The plane-wave kinetic energy cutoff is set to 420~eV. The convergence criteria for the total energy and atomic force are $10^{-4}$~eV and 0.01~eV/\r{A}, respectively. The Heyd-Scuseria-Ernzerhof HSE06 hybrid functional~\cite{heyd_hybrid_2003} is employed to calculate the nuclear quadrupole interaction constant. 
For the \textsuperscript{17}O nucleus (with nuclear spin $I = 5/2$) in OV defects, the nuclear quadrupole interaction constant is defined as 
\begin{equation}
\mathrm{C_{q}} = eQ_{\mathrm{O}}V_{zz}/h \text{,}
\end{equation}
where $Q_{\mathrm{O}}$, $e$, $V_{zz}$, and $h$ denote the nuclear electric quadrupole moment, elementary charge, the principal component of electric field gradient (EFG) at the nucleus, and Planck's constant, respectively. By using $Q_{\mathrm{O}} = -25.58 \times 10^{-31}\,\mathrm{m^2}$ ~\cite{pyykko_year-2008_2008}, the quadrupole coupling constants are calculated to be $\mathrm{C_{q}} = 7.04$ MHz and $\mathrm{C_{q}} = 6.83$ MHz for OV($hh$) and OV($kk$), respectively. For OV($kh$) and OV($hk$) configurations, the corresponding values are $\mathrm{C_{q}} = 7.15$ MHz and $\mathrm{C_{q}} = 6.89$ MHz, respectively. In these low-symmetry configurations, the asymmetry parameters defined as 
\begin{equation}
\eta = (V_{yy} - V_{xx})/V_{zz} 
\end{equation}
are at 0.01 and 0.04, respectively. We note that the quadrupole moment of $^{17}$O does not contribute to the splitting in the respective ODMR spectra.

\subsection{Hyperfine coupling of PL6 with $^{29}$Si and $^{13}$C}

\begin{figure*}[htbp] 
\centering 
\includegraphics[scale=0.7]{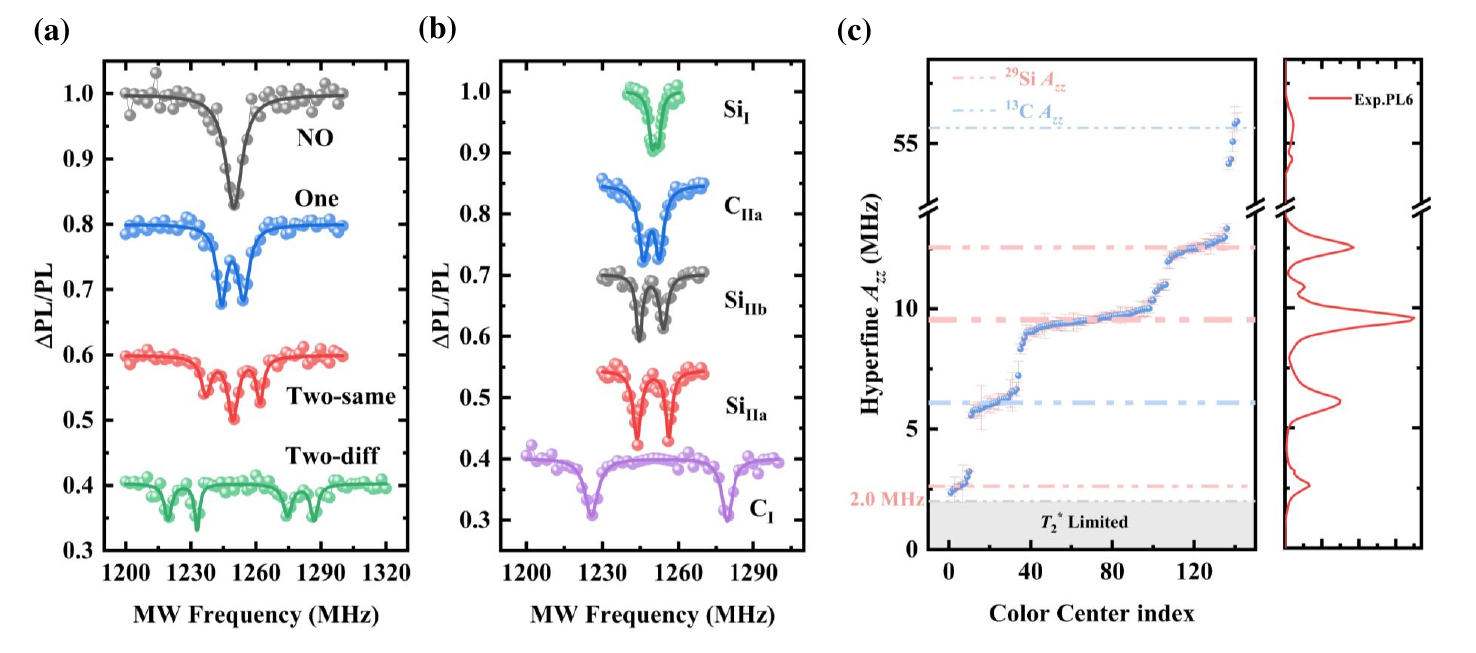} 
\caption{ 
{Hyperfine $A_{zz}$ parameters for $^{29}$Si and $^{13}$C coupled to PL6 at $\gamma B_z = 101.0 \pm 0.3$ MHz. 
(a) Representative multi-peak hyperfine splitting patterns arising from coupled nuclear spins. 
(b) ODMR spectra for specific coupling configurations. 
(c) Distribution summary of the PL6 hyperfine coupling parameters.}}
\label{S14} 
\end{figure*}

To further verify the microscopic structure of PL6, we statistically analyzed its hyperfine couplings to nearby $^{29}$Si and $^{13}$C nuclear spins and compared the extracted parameters with the theoretical O$_\text{C}$V$_\text{Si}$($hh$) model. In total, 141 PL6 defects exhibiting strong hyperfine coupling to $^{29}$Si or $^{13}$C were analyzed. Figure~S\ref{S14}a presents several representative multi-peak hyperfine patterns arising from nearby nuclear spins, where the ODMR contrast decreases as the number of resolved coupled peaks increases. Figure~S\ref{S14}b shows typical coupling configurations, including the Si$_\text{I}$, Si$_\text{IIa}$, Si$_\text{IIb}$, C$_\text{I}$, and C$_\text{IIa}$ sites. By extracting the corresponding $A_{zz}$ parameters and broadening them with Gaussian peaks, we obtained the distributions shown in Fig.~S\ref{S14}c. The experimental results are in good agreement with the theoretical predictions of the O$_\text{C}$V$_\text{Si}$($hh$) model~\cite{zhao2025analyzing}. This consistency provides further support for assigning PL6 to the O$_\text{C}$V$_\text{Si}$($hh$) center and complements the direct structural identification based on the $^{17}$O hyperfine measurements.

\subsection{Annealing under argon and vacuum atmospheres}
\begin{figure*}[htbp] 
\centering 
\includegraphics[scale=0.4]{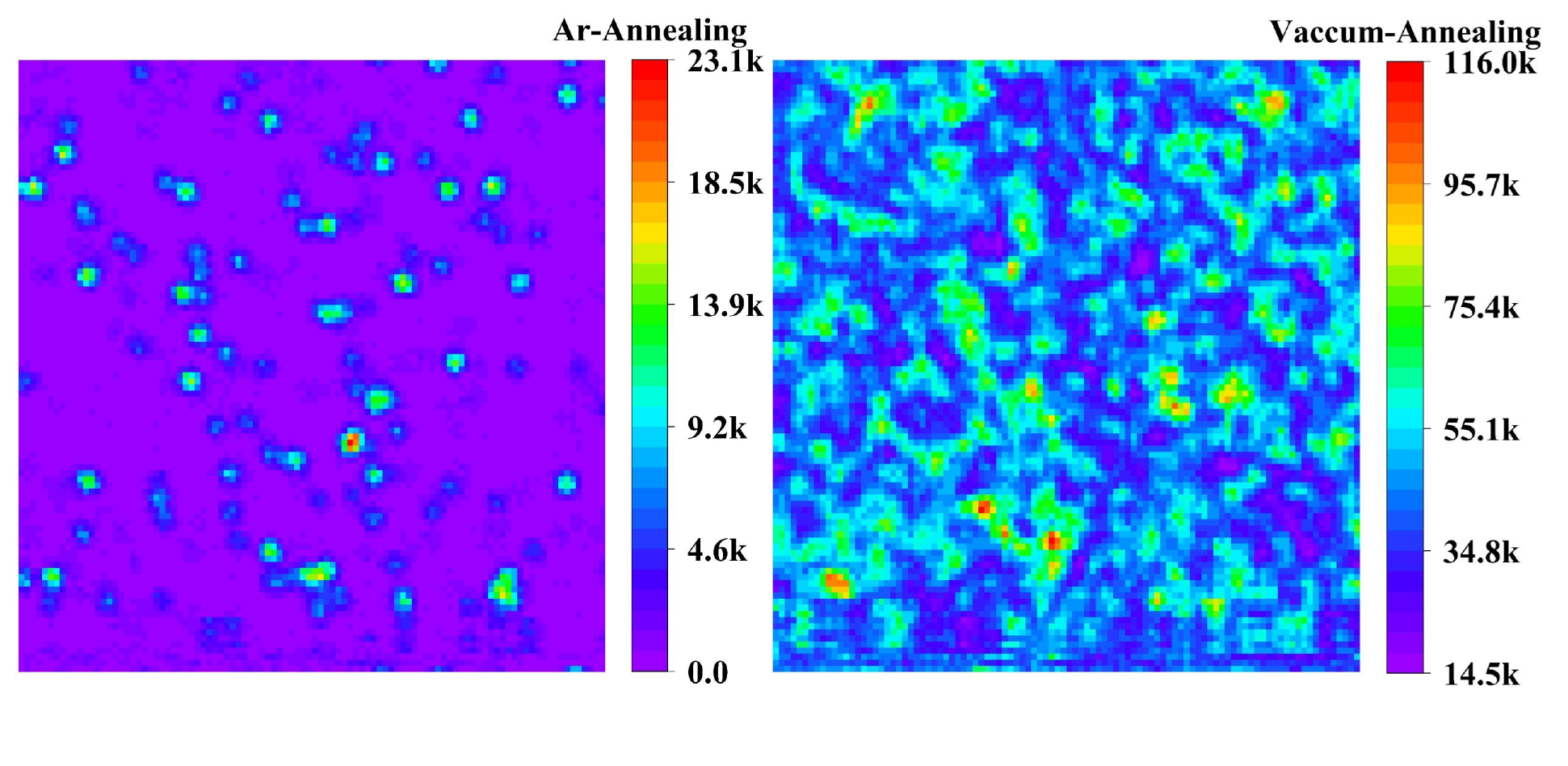} 
\caption{{Fluorescence scanning images of samples implanted at a dose of $1\times10^8$ cm$^{-2}$ and annealed under argon or vacuum at 1050$^\circ$C for 0.5 h.}} 
\label{S15} 
\end{figure*}
We compared annealing under argon and vacuum atmospheres to tune the concentration of low-density color centers. As shown in Fig.~S\ref{S15}, annealing in argon yields cleaner isolated emission spots, whereas vacuum annealing results in a higher color center density. One possible explanation is that argon provides a protective environment during annealing, which could reduce the formation of oxygen-related modified divacancies. Therefore, in the engineering of oxygen-related modified divacancy color centers, the defect concentration can be conveniently adjusted by selecting the annealing atmosphere according to the specific experimental requirements.\\
\\
\\